\documentclass[twocolumn,tighten]{aastex61}

\usepackage{gensymb}
\usepackage{rotating,epsfig}
\usepackage{natbib}
\def\msun{\rm M_{\sun}}

\def\um{$\mu$m }

\submitjournal{ApJ}

\shorttitle{\emph{Herschel} PACS observations in Ori OB1}
\shortauthors{Mauc\'o et al.}

\begin{document}
\title{Herschel PACS observations of 4-10 Myr old Classical T Tauri stars in Orion OB1}

\author[0000-0001-8284-4343]{Karina Mauc\'o}
\affiliation{Instituto de Radioastronom\'ia y Astrof\'isica, UNAM, Campus Morelia,
Antigua Carretera a P\'atzcuaro No. 8701, Col. Ex Hacienda San José de la Huerta,
Morelia, Michoac\'an, C.P. 58089, M\'exico}
\email{k.mauco@irya.unam.mx}

\author[0000-0001-7124-4094]{C\'esar Brice\~{n}o}
\affiliation{Observatorio Interamericano Cerro Tololo, AURA/CTIO, Casilla 603, La Serena, Chile}

\author[0000-0002-3950-5386]{Nuria Calvet}
\affiliation{Department of Astronomy, University of Michigan, 311 West Hall, 1085 South University Avenue, Ann Arbor, MI 48109, USA}

\author[0000-0001-9797-5661]{Jes\'us Hern\'andez}
\affiliation{Instituto de Astronom\'ia, UNAM, Campus Ensenada, 
Carretera Tijuana-Ensenada km 103, 22860 Ensenada, B.C. M\'exico}

\author[0000-0002-1081-9445]{Javier Ballesteros-Paredes},
\affiliation{Instituto de Radioastronom\'ia y Astrof\'isica, UNAM, Campus Morelia,
Antigua Carretera a P\'atzcuaro No. 8701, Col. Ex Hacienda San José de la Huerta,
Morelia, Michoac\'an, C.P. 58089, M\'exico}

\author{Omaira Gonz\'alez}
\affiliation{Instituto de Radioastronom\'ia y Astrof\'isica, UNAM, Campus Morelia,
Antigua Carretera a P\'atzcuaro No. 8701, Col. Ex Hacienda San José de la Huerta,
Morelia, Michoac\'an, C.P. 58089, M\'exico}

\author{Catherine Espaillat}
\affiliation{Department of Astronomy, Boston University, 725 Commonwealth Avenue Boston, MA 02215}

\author[0000-0001-7318-6318]{Dan Li}
\affiliation{Department of Physics \& Astronomy, University of Pennsylvania, Philadelphia, PA 19104, USA}

\author{Charles M. Telesco}
\affiliation{Department of Astronomy University of Florida Gainesville, FL 3261-2055 USA}

\author[0000-0001-6559-2578]{Juan Jos\'e Downes}
\affiliation{Centro de Investigaciones de Astronom\'ia, CIDA, Av. Alberto Carnevalli, Sector ``La Hechicera", M\'erida, Venezuela.
Apartado Postal 264, M\'erida 5101-A, M\'erida - Venezuela.}

\author[0000-0003-1283-6262]{Enrique Mac\'ias}
\affiliation{Department of Astronomy, Boston University, 725 Commonwealth Avenue Boston, MA 02215}

\author[0000-0001-8642-1786]{Chunhua Qi}
\affiliation{Harvard-Smithsonian Center for Astrophysics 60 Garden Street, Mail Stop 42 Cambridge, MA 02138}

\author[0000-0003-1263-808X]{Ra\'ul Michel}
\affiliation{Instituto de Astronom\'ia, UNAM, Campus Ensenada, 
Carretera Tijuana-Ensenada km 103, 22860 Ensenada, B.C. M\'exico}

\author{Paola D'Alessio}
\altaffiliation{author is deceased}
\affiliation{Instituto de Radioastronom{\'\i}a y Astrof{\'\i}sica, UNAM, Campus Morelia,
Antigua Carretera a P\'atzcuaro No. 8701, Col. Ex Hacienda San José de la Huerta,
Morelia, Michoac\'an, C.P. 58089, M\'exico}

\author{Babar Ali}
\affiliation{Space Sciences Institute, Boulder, CO}

\begin{abstract}
We present \emph{Herschel} PACS observations of 8 Classical T Tauri Stars in the $\sim 7-10$ Myr old OB1a and the $\sim 4-5$ Myr old OB1b Orion sub-asscociations. 
Detailed modeling of the broadband spectral energy distributions,
particularly the strong silicate emission at 10 $\mu$m,
shows that these objects are (pre)transitional disks with some amount of small optically thin dust inside their cavities, ranging from $\sim 4$ AU to
$\sim 90$ AU in size. We analyzed \emph{Spitzer} IRS spectra for 
two objects in the sample: CVSO-107 and CVSO-109.
The IRS spectrum of CVSO-107 indicates
the presence of crystalline material inside its gap while 
the silicate feature of CVSO-109 is characterized 
by a pristine profile produced by amorphous silicates;
the mechanisms creating the optically thin dust 
seem to depend on disk local conditions.
Using millimeter photometry 
we estimated dust disk masses for CVSO-107 and CVSO-109
lower than the minimum mass of solids needed
to form the planets in our Solar System, 
which suggests that giant planet formation
should be over in these disks.  
We speculate that the presence and maintenance of optically thick material in the inner regions of these
pre-transitional disks might point to low-mass planet formation.    
 
\end{abstract}


\section{Introduction}
\label{sec:intro}

Understanding how solid material in protoplanetary disks evolves
from conditions similar to those in the Interstellar Medium (ISM)
to planetary embryos and beyond,
requires both theoretical developments and observational
constraints. As many complex processes are at play, observations
are essential to inform theory and set constraints on the
multiple effects occurring in these disks.

Many studies have now revealed stars with inner disks devoid of 
optically thick material
$-$the so-called transitional disks (TD)$-$ and with spectral energy
distributions (SEDs) characterized by
small or negligible near-infrared excesses but significant emission 
in the mid-infrared and beyond 
\citep{strom89,skrutskie90,calvet02,espaillat07,espaillat08a,espaillat14}. 
This morphology has been interpreted as cavities in the inner regions
and has been confirmed through (sub)millimeter interferometric imaging
and, recently, by 
spatially resolved images 
\citep[e.g.,][]{hughes09,brown09,andrews09,isella10,andrews11b,andrews11a,van_Dishoeck15,carrasco16},
especially those taken with the Very Large Telescope (VLT)/SPHERE and the Atacama Large Millimeter/submillimeter Array (ALMA)
\citep{ALMA15,nomura16,schwarz16,andrews16,vanBoekel17}. 
A subset of these disks, commonly called pre-transitional disks (PTD), 
show similar features but with substantial near-infrared excesses over the 
stellar photosphere.
This excess has been explained by 
an optically thick disk located close to the star,
separated from the outer disk by a gap 
\citep{espaillat08a,espaillat10,espaillat11}. 

Transitional and pre-transitional
disks
are thought to be at
an important phase of disk evolution. 
The distinct SEDs of these sources have 
puzzled researchers over the years.  
For instance, dust clearing mechanisms in TD/PTD are still under debate. 
It is uncertain if multiple processes act simultaneously or
a single process dominates the evolution.
Observations made with the \emph{Spitzer Space Telescope} \citep{werner04} have been widely used
to identify TD/PTD and
to characterize their IR emission.
Moreover, \emph{Spitzer} IRS spectra have also provided unprecedented
details regarding disk cavities and dust properties
within them.
Extensive modeling of several TD/PTD around
T Tauri stars (TTS) has been made
\citep{calvet02,calvet05,uchida04,dalessio05,espaillat07,espaillat08b,espaillat10,mcclure10,mcclure12,mcclure13b}.
Some mechanisms have been proposed to explain the substantial dust clearing observed on these disks, e.g. grain growth, 
photoevaporation
and interaction with embedded planets or stellar companions
\citep{espaillat14}.
Many researchers have proposed planet formation as the most
likely mechanism, since
models of planet-disk interaction have resulted in cleared disk regions
\citep[e.g.,][]{paardekooper04,zhu11,dodson-Robinson11,dipierro17,dong17}.
Additionally, thanks to new generation-high angular resolution instruments, 
observational evidence  
implying the presence of (proto)planets has increased over the years
\citep[e.g.,][]{huelamo11,kraus11,pinilla15,follette15,sallum15,sallum16,dejuanOvelar16,andrews16}.

Studies of disk frequencies as a function of age
have set timescales for disk evolution of $\sim$5-7 Myr for late-type (K to M) stars \citep[e.g.,][]{hernandez07b}.
However, most disk studies have concentrated on populations $\lesssim $ 2 Myr (i.e., Taurus,
the Orion Nebula Cluster), and much less information exists for
older populations, especially around 10 Myr. The main reason is that already at ages $\sim 4$ Myr
the parent molecular clouds have largely dissipated, such that these somewhat older stars are harder to
identify among the general field population. It is not surprising that many of 
the $\sim 4 - 10$ Myr old groups have been discovered in the last $\sim 20$ years.
The TW Hya association \citep{webb1999} and the $\eta$ Cha
cluster \citep{mamajek1999} are among the nearest 10 Myr old groups, but contain only around 20 stars.
The Scorpius-Centaurus OB association, with ages $\sim 5-20$ Myr
\citep[e.g.][]{preibisch2008,pecaut_mamajek2016} is the closest OB association ($\sim 130$ pc), and has been studied extensively as a source of older PMS disk-bearing stars. The Upper-Scorpius (US) region has $\sim 800$ members reported by \cite{luhman12}, though spectroscopic confirmation of many low-mass members is still ongoing \citep{pecaut_mamajek2016};
the age is still debated, proposed to be in the range 4-10 Myr \citep{preibisch2008,pecaut_mamajek2016}.
ALMA submm studies in US have reported dust properties and disk sizes in samples of $\sim 100$ disk systems
\citep{carpenter06,carpenter09,barenfeld16,barenfeld17}.

As the closest region with active low and high-mass star formation (d$\sim 400$ pc),
the Orion OB1 association contains large samples of young stars, spanning ages from the 
protostellar stage up to ``older" PMS stars, and sharing a common origin \citep{bally2008,briceno2008}.
While Sco-Cen only hosts slightly more evolved stars ($\ga 4$ Myr) and no recognizable star clusters, Orion has populous stellar aggregates at all ages up to $\sim 10$ Myr. 
These young, dense stellar groups provide an opportunity for exploring the evolution of protoplanetary disks in clustered environments.
At the young end of optically visible PMS stars, the $\sim 1$ Myr old Trapezium cluster contains 
$\ga 2000$ stars \citep{muench2008}, 
and the $\sim 3$ Myr $\sigma$ Ori cluster has over 300 confirmed members \citep{hernandez14}. 
At the ``old" end of the PMS age spectrum, the $\sim 10$ Myr old 25 Ori cluster \citep{briceno07b} has $\sim 250$ members, 
which have been characterized spectroscopically and photometrically by us in a consistent and uniform way.
Here we present Herschel Space Telescope $70$ $\mu$m and $160$ $\mu$m observations of four fields in the Orion OB1
association, targeting a limited subset of the stellar population in the 4-10 Myr age range, including the 25 Ori cluster.

Disk fluxes are strongly dependent on the stage of dust
evolution at the wavelength range probed by Herschel. WISE covered wavelengths from $3.6$ $\mu$m to $22$ $\mu$m,
and Spitzer was most sensitive from $3.6$ $\mu$m to $24$ $\mu$m, with limited sensitivity at $70$ $\mu$m.
Therefore, studies of dust depletion have mostly been limited to the inner disk regions.
With Herschel we now have a window into longer wavelengths important for disk studies.
Additionally,
the smaller beam size and higher spatial resolution
of PACS compared with {\it Spitzer}/MIPS results in a lower rate
of confusion with background sources making it easier for Herschel to detect faint sources.
Combining the Herschel data with optical V, R, I, near infrared J, H, K photometry, 
and mid-IR data from Spitzer/WISE, we assemble spectral energy distributions
(SED) for 8 disk-bearing sources, that we then fit with detailed irradiated accretion disk models
to infer the structure, characteristics and evolutionary state of these disks.
In \S \ref{sec:obsred} we discuss observations, sample selection and data reduction;
our analysis and results are shown in \S \ref{sec:results} where we
present the observed SEDs of our objects (\S \ref{sec:seds}), the stellar parameters and mass accretion rates
estimates (\S \ref{sec:stellar_para}), a description of our disk models (\S \ref{sec:disk models})
and the method we used to model the SEDs (\S \ref{sec:modeling}); 
the main 
results are discussed in \S \ref{sec:discussion}
and, finally, our conclusions are listed in \S \ref{sec:conclusion}.

\section{Observations}
\label{sec:obsred}

In this section we summarize the observational data sets obtained for our sample. Optical photometry was used to characterize the stellar properties, which are inputs in our models, and as an indicator of how variable these stars are, which was included as an additional uncertainty in the $\chi^{2}$ estimate. Mid-IR data was used in the modeling of the silicate bands, particularly at 10 $\mu$m. PACS photometry allowed the characterization of the outer disk edge, while sub-mm data was used to estimate disk mass and to constrain the properties of the outer disk. 

\subsection{Sample}
\begin{figure}
\centering
\includegraphics[angle=270,scale=0.4]{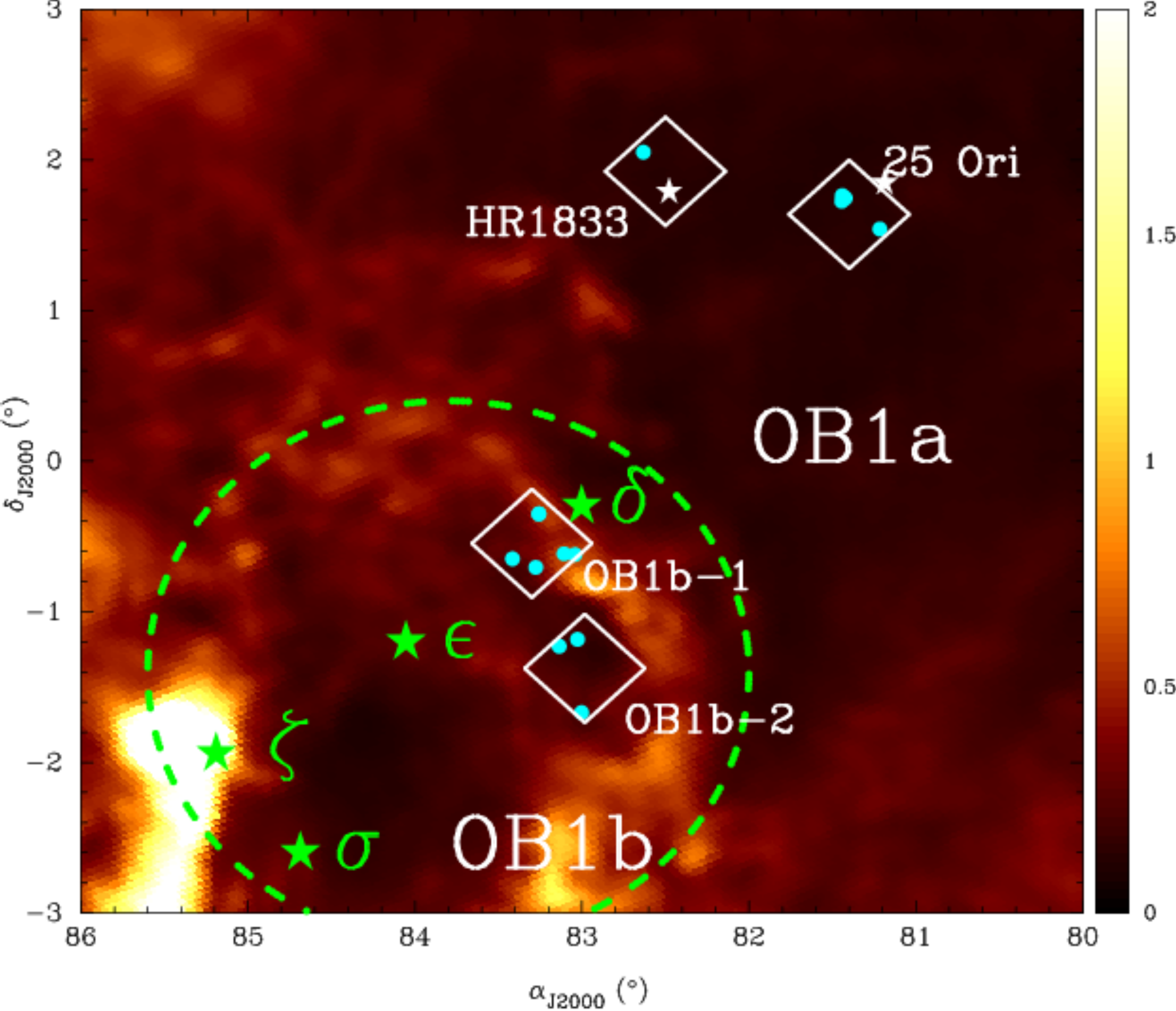}
\caption{Location of the four \emph{Herschel} PACS fields in the Orion OB1 association,
overlayed on the extinction map from \cite{schlegel98}. The Orion OB1b region encompasses 
the three Orion belt stars, and is delimited by
the large dashed lined circle \citep{briceno05}; the OB1a region 
is located west and north of OB1b, 
and includes
the 25 Ori cluster and the HR 1833
stellar aggregate. 
The TTS detected at 70 $\mu$m in the PACS fields are shown as cyan dots.
Our Herschel fields are all in regions of low extinction with $A_V \lesssim 1$ mag.
}
\label{pacs_fields}
\end{figure}

We targeted a set of $165$ TTS distributed in four fields
(Figure \ref{pacs_fields}),
two 
in the $\sim 4-5$ Myr old Orion OB1b subassociation 
and two in the $\sim 7-10$ Myr old Orion OB1a region \citep{briceno05}. 
These stars have been confirmed as members of the OB association based on their
K and M spectral types, H$\alpha$ emission, and the presence of Li~I ($\lambda\, 6707$\AA) in absorption
\citep{briceno05}.
For this sample, we also have multi-band, multi-epoch optical photometry
from the CVSO \citep[][Brice\~no et al. 2018, in preparation]{briceno05} and from the 
Sloan Digital Sky Survey (SDSS)\footnote{http://www.sdss3.org/dr9/}, 
near infrared J, H, and $\rm K_S$ magnitudes from the Two Micron All Sky Survey \citep[2MASS,][]{skrutskie06}, Z, J, H, and $\rm K_S$ photometry from the Visible and Infrared Survey Telescope for Astronomy Science Verification Survey of Orion OB1 \citep[VISTA,][]{petr-gotzens2011}, 
infrared photometry from the Infrared Array Camera \citep[IRAC,][]{fazio04} and the Multiband Imaging Photometer for Spitzer \citep[MIPS,][]{rieke04} at 3.6, 4.5, 5.8, 8.0, and 24.0 $\mu$m from our GO-13437 and GO50360 programs \citep[][Brice\~no et al. 2018, in preparation]{hernandez07a}, 
and at 3.6, 4.5, 6, 12, and 22 $\mu$m from the Wide-field Infrared Survey Explorer \citep[WISE,][]{wright2010}. 
The two fields in the Ori OB1a region are roughly centered on the 25 Ori cluster 
\citep{briceno07b} and the HR1833 stellar aggregate (Brice\~no et al. 2018, in preparation),
and combined encompass $118$ confirmed TTS. The two Ori OB1b fields
contain $47$ confirmed TTS. 
The spectral types for the young stars in both regions are similar,
and span the range K4 to M5, which at the ages of our stars
corresponds to masses $0.12 \lesssim M/M_{\odot} \lesssim 1.2$ \citep{baraffe98}.

\subsection{Herschel PACS photometry}

Our Herschel/PACS imaging survey of the Orion OB1 fields was carried out 
with 8 unique Herschel observations at $70$ $\mu$m and $160$ $\mu$m, obtained on March 16, 18, 28, and 29, 2012.  
We used the ``scan map" observational 
template with medium scan speed ($20\arcsec$/s) to map a square field 
$30\arcmin$ per side.  Each scan line was $30\arcmin$ long, and 134 overlapping 
scan lines with a stepsize of 15 arcseconds were sufficient to reach the 
target size and sensitivity. Each field was observed twice at orthogonal scan directions.  
This technique is commonly used to mitigate the low-frequency drift of the 
bolometer timelines.  With this configuration, we aimed at reaching 
a 1-$\sigma$ point source sensitivity of 2.6 mJy and 6 mJy in the 
blue and red channels, respectively.    
Our Herschel program identifier is $\rm OT1\_ncalvet\_1$.

We rely on the Herschel data processing pipelines \citep{ott2010} for the initial data
processing and begin further processing at the so-called Level 1 stage. 
All data discussed here are based on the ``FM6" version of the PACS calibration \citep{balog2013} and
processed with version 9 of the Herschel Interactive Processing Environment
\citep[HIPE][]{ott2010} software.

We processed the Level 1 data with the {\sl Scanamorphos} technique, a map-making software
developed and described by \cite{roussel2012}.
Scanamorphos removes the 1/f noise \footnote{The term 1/f noise is used here to generically describe bolometer signal drifts that are inversely correlated with their Fourier frequency.} by making use of the redundancy built in the observations.
Readers are referred to \cite{roussel2012} for details about the processing steps. 
Scanamorphos preserves astrophysical emission on all spatial scales,
ranging from point sources to extended structures with scales just below the map size;
therefore, the maps produced are suitable for both spatially extended and point sources.

We performed source detection on the $70$ $\mu$m images processed with Scanamorphos, 
using the \texttt{daofind} task in IRAF\footnote{IRAF is distributed by the National Optical Astronomy Observatory, which is operated by the Association of Universities for Research in Astronomy (AURA) under a cooperative agreement with the National Science Foundation.}.
We then proceeded to obtain aperture photometry on both $70$ $\mu$m and $160$ $\mu$m channels,
using the IRAF \texttt{apphot} task.
Following \cite{fischer2013},
 for the $70$ $\mu$m images we used an aperture radius of $9.6\arcsec$, inner sky annulus
 radius of $9.6\arcsec$ and sky annulus width of $9.6\arcsec$;
 for the 160 $\mu$m images we used an aperture radius of $12.8\arcsec$, inner sky annulus
 radius of $12.8\arcsec$ and a $12.8\arcsec$ sky annulus width.
 Because the pixel scale is 1$\arcsec$/pixel for the 70 $\mu$m images and
 $2\arcsec$/pixel for the 160 $\mu$m images, these apertures correspond to
 9.6 pixels and 6.4 pixels respectively.
 Though the relatively small apertures require large aperture corrections:
 0.7331 for the blue channel ($70$ $\mu$m) and 0.6602 for the red channel ($160$ $\mu$m),
 they minimize uncertainties due to large variations
 in the sky background in regions with significant nebulosity, especially with the $160$ $\mu$m 
 images.
 Photometric errors were determined as the sum in quadrature of the measurement error
 and the calibration error. 

Of the 165 TTS located within the PACS fields, only 16 are classified as
disk sources based on their $\rm K_s - [4.5]$ excess, using the criterion 
in Figure 1 of \cite{luhman12}.
The number of disk sources goes up to 25 if we use the excess emission at
$\rm K_s - [8.0]$ or $\rm K_s - [24]$.
We detected 8 of these disk sources with PACS (33-50\% depending on the disk indicator),
all classified as accreting Classical TTS (CTTS)\footnote{CTTS are T Tauri stars still actively accreting from a circumstellar disk; they are usually classified as such from low-resolution spectra showing a strong H$\alpha$ emission line, with an equivalent width above the value expected for chromospheric emission at the spectral type of the star (\cite{white03}).} based on our optical spectra. These objects
are also tagged as Class II stars based on the IRAC SED slopes in
\cite{hernandez07a}.
They are distributed as follows: 2 are located in the OB1a region, specifically 
in the 25 Ori cluster and the HR 1833 group
(Brice\~no et al. 2018, in preparation), and 6 in the two OB1b PACS fields.
We present Herschel PACS photometry at $70$ $\mu$m and $160$ $\mu$m for these
objects.
PACS photometry is shown in Table~\ref{tab:observations} in the following order: target CVSO ID,
2MASS ID, right ascension, declination, $70$ $\mu$m flux, $160$ $\mu$m flux, and the 
location of each source.

\begin{deluxetable*}{lcccccc}

\tablecaption{PACS Photometry for CTTS in the OB1a and OB1b Subassociations \label{tab:observations}}
\tablehead{
\multicolumn{1}{c}{CVSO ID} & \multicolumn{1}{c}{2MASS ID} & \multicolumn{1}{c}{RA (J2000.0)} & \multicolumn{1}{c}{DEC (J2000.0)} & \multicolumn{1}{c}{F$_{70}$} & \multicolumn{1}{c}{F$_{160}$} & \multicolumn{1}{c}{location} \\ \noalign{\smallskip}
\colhead{} & \colhead{}  & \colhead{(hh:mm:ss)} & \colhead{(hh:mm:ss)} & \colhead{(mJy)} 
& \colhead{(mJy)} & \colhead{} 
}
\startdata
CVSO-35     & 05254589+0145500  &  05:25:45.90  & +01:45:50.3  & 18.83   $\pm$ 1.19   & 12.77  $\pm$ 2.48  & 25 Ori \\
CVSO-104    & 05320638-0111000  &  05:32:06.45  & -01:11:00.3  & 77.41   $\pm$ 2.35   & 56.28  $\pm$ 3.78  & OB1b   \\
CVSO-107    & 05322578-0036533  &  05:32:25.77  & -00:36:53.2  & 105.34  $\pm$ 3.32   & 106.51 $\pm$ 8.34  & OB1b   \\
CVSO-109    & 05323265-0113461  &  05:32:32.66  & -01:13:46.0  & 69.38   $\pm$ 2.69   & 38.97  $\pm$ 5.36  & OB1b   \\
CVSO-114NE  & 05330196-0020593  &  05:33:01.97  & -00:20:59.3  & 30.67   $\pm$ 2.56   & 94.33  $\pm$ 12.12 & OB1b   \\
CVSO-121    & 05333982-0038541  &  05:33:39.82  & -00:38:53.9  & 71.59   $\pm$ 2.26   & 47.29  $\pm$ 3.11  & OB1b   \\
CVSO-238    & 05320040-0140110  &  05:32:00.40  & -01:40:11.0  & 8.126   $\pm$ 1.22   & 5.13   $\pm$ 2.92  & OB1b   \\   
CVSO-1265   & 05303164+0203051  &  05:30:31.66  & +02:03:05.2  & 35.32   $\pm$ 1.45   & 48.56  $\pm$ 2.95  & HR1833 \\
\enddata
\end{deluxetable*}

\subsection{CanariCam Photometry}\label{sec:cana_phot}

We observed 5 sources in the OB1 fields (2 in OB1a: CVSO-35, CVSO-1265 and 3 in OB1b: CVSO-104, CVSO-114NE and CVSO-121) in the 
Si2 (8.7 $\mu$m), Si4 (10.3 $\mu$m), Si5 (11.6 $\mu$m), and Si6 (12.5 $\mu$m) narrow band silicate filters
on 2014 September 22 and 23 and October 3 and 5 with the CanariCam\footnote{http://www.gtc.iac.es/instruments/canaricam/canaricam.php} instrument \citep{telesco03}
on the Gran Telescopio de Canarias (GTC). 
The selection criteria was targets with \emph{Spitzer}/IRAC and \emph{Herschel}/PACS (70 $\mu$m and 160 $\mu$m) detections. 
CanariCam has a 26\arcsec x 19\arcsec field of view
with a detector plate scale of 0\farcs08 per pixel.
The reduction of the data was done using the CanariCam data reduction
pipeline (RedCan). RedCan produces flux-calibrated images using
the associated standard star images along with their theoretical spectra reported by \citet{cohen99}. An 
extensive description of the RedCan pipeline can be found in \citet{omaira13}. 
Table ~\ref{tab:cana_table} summarizes the CanariCam photometry. 
Column 1 shows the CVSO ID following \citet{briceno05} while Columns 2-5 indicate 
the fluxes in the four narrow band silicate filters for each source. Errors are about 15\% of the photometric value in each band \citep{alonso16}.

\begin{deluxetable}{lcccc}

\tablecaption{CanariCam photometry of 5 PACS sources in the OB1a and OB1b subassociations \label{tab:cana_table}}
\tablehead{
\multicolumn{1}{c}{CVSO ID}  & \multicolumn{1}{c}{$F_{\rm Si2}$} & \multicolumn{1}{c}{$F_{\rm Si4}$} & \multicolumn{1}{c}{$F_{\rm Si5}$} & \multicolumn{1}{c}{$F_{\rm Si6}$}\\
\noalign{\smallskip}
\colhead{} & \colhead{(Jy)}  & \colhead{(Jy)} & \colhead{(Jy)} & \colhead{(Jy)}
}
\startdata
CVSO-35     & 68.86  & 106.96 & 120.58   & 121.09\\
CVSO-104    & 25.08  & 50.75  & 48.28    & 16.92 \\
CVSO-114NE  & 29.01  & 23.97  & 24.39    & 25.48 \\
CVSO-121    & 35.58  & 50.22  & 35.11    & 17.05 \\
CVSO-1265   & 27.51  & 56.73  & 34.98    & 20.70 \\
\enddata
\tablecomments{Column 1: CVSO ID following \citet{briceno05}; Column 2: Si2 flux;
Column 3: Si4 flux; Column 4: Si5 flux; Column 5: Si6 flux. Errors are about 15\% the photometric value in each band.
}
\end{deluxetable}
 
\subsection{Optical Photometry}
We obtained new optical photometry of a subset of our stars for which there was no existing SDSS photometry, or stars with UV excesses, or those with strong photometric variability between the CVSO and the SDSS photometric bands. These observations were obtained at the 4.3m Discovery Channel telescope at Lowell Observatory, Arizona, USA, and the 0.84m telescope at the San Pedro M\'artir National Astronomical Observatory in Baja California, M\'exico.\\\\

\subsubsection{The Discovery Channel Telescope}
We observed CVSO-107 and CVSO-109
in the U, B, V, R, and I Johnson-Cousins filters on 2013 November 29 with the Large Monolithic Imager (LMI)\footnote{http://www2.lowell.edu/rsch/LMI/specs.html} on the Discovery Channel Telescope (DCT).
The LMI has a 12\farcm5 x 12\farcm5 field of view, with an unbinned pixel size of 0\farcs12; we utilized the 2 x 2 pixel binning mode, resulting in a pixel scale of 0\farcs24 per pixel. 
We used IRAF to carry out bias and flat-field corrections, using twilight flats, and then to derive aperture photometry interactively. 
The optical photometry is listed in Table \ref{tab:optphot} where we show the CVSO ID following \citet{briceno05} in Column 1 and 
in Column 2-6 we present the photometry of the U, B, V, $\rm R_c$, and $\rm I_c$ filters of our sample.

\begin{deluxetable*}{lccccc}

\tablecaption{Optical Photometry\label{tab:optphot}}
\tablehead{
\colhead{CVSO ID} & \colhead{U} & \colhead{B} & \colhead{V} & \colhead{R$_{c}$} & \colhead{I$_{c}$}
}
\startdata
CVSO-35$^{b}$   &  16.744$\pm$0.007 & 15.578$\pm$0.002 & 14.059$\pm$0.002 & 13.226$\pm$0.002 & 12.388$\pm$0.002 \\
CVSO-104$^{b}$  &  15.973$\pm$0.017 & 15.659$\pm$0.015 & 15.026$\pm$0.076 & 14.442$\pm$0.034 & 13.319$\pm$0.007 \\
CVSO-107$^{a}$  & 15.287$\pm$0.005 & 15.548$\pm$0.004 & 14.628$\pm$0.002 & 13.788$\pm$0.004 & 12.831$\pm$0.003 \\
CVSO-109$^{a}$  & 14.897$\pm$0.012 & 14.997$\pm$0.005 & 14.044$\pm$0.003 & 13.260$\pm$0.006 & 12.219$\pm$0.005 \\
CVSO-114NE$^{b}$&  16.592$\pm$0.005 & 16.984$\pm$0.004 & 15.986$\pm$0.010 & 15.086$\pm$0.006 & 13.752$\pm$0.007 \\
CVSO-121$^{b}$  &  15.886$\pm$0.004 & 15.598$\pm$0.001 & 14.392$\pm$0.006 & 13.528$\pm$0.003 & 12.672$\pm$0.002 \\
\enddata
\tablecomments{$^{a}$ DCT photometry. $^{b}$ SPM photometry. 
}
\end{deluxetable*}

\subsubsection{San Pedro M\'artir}

CVSO-35, CVSO-104, CVSO-114NE, and CVSO-121 were observed in the
UBV$(\rm RI)_c$ system, during an open cluster campaign, in December 4 and 5
2016 at the San Pedro Martir (SPM) Observatory with the 0.84-m telescope and
the \textit{Marconi 3} CCD detector (a deep depletion e2v CCD42-40
chip with gain of 1.83 e$^-$/ADU and readout noise of 4.7 e$^-$). 

The
field of view was $7.4^{\prime}\times7.4^{\prime}$ and binning
2$\times$2 was used during the observations. In order to properly
measure both bright and dim stars, different exposure times were
employed. We used 2, 20, and 200s in both R and I filters, 4, 40, and 400s
for the V filter, 6, 60, and 600s for the B filter, and 10, 100, and 1000s for
the U filter. Standard stars in Landolt fields were observed during 
the night in order to calibrate the photometry. 
The data reduction was done with IRAF, following the standard procedure
for correcting bias and flat field frames. Instrumental magnitudes were
derived using standard PSF photometry. Transformation
equations, based on the observed standard stars, were then applied to
convert the instrumental magnitudes to calibrated magnitudes. The
resulting photometry is shown in Table~\ref{tab:optphot}. 

\subsection{Sub-mm Photometry}
We observed CVSO-107 and CVSO-109 on 2010 January 1 with the Submillimeter Array (SMA) on top of Mauna Kea, HI, using the compact array configuration (projected baselines of 9.8--81.2 m). The weather was excellent with the 225 GHz opacity around 0.05 and stable atmospheric phase. The double sideband system temperatures were 72--156 K. Calibration of the visibility phases and amplitudes was achieved with observations of the quasar J0532+075, at intervals of about 30 minutes. The bandpass response was calibrated using 3C454.3. Observations of Uranus provided the absolute scale for the flux density calibration and the derived flux of J0532+075 was 0.71 Jy. The data were calibrated using the MIR software package\footnote{http://www.cfa.harvard.edu/$\sim$cqi/mircook.html}. Continuum images were generated and CLEANed using standard techniques in the MIRIAD software package. Fluxes are listed in Column 2 of Table \ref{tbl:redphot}.\\\\

\begin{deluxetable}{lccc}

\tablewidth{0pt}
\tablecaption{Sub-mm Fluxes\label{tbl:redphot}}
\tablehead{
\colhead{CVSO ID} & \colhead{F$_{1300}$ (mJy)}
}
\startdata
CVSO-107  & 7.1$\pm$1.2 \\
CVSO-109  & 3.2$\pm$1.3 \\
\enddata
\end{deluxetable}

\subsection{Spitzer IRS Spectra}

CVSO-107 and CVSO-109 were observed by the \emph{Spitzer} IRS
instrument on 2006 March 18 (AORKEY: 14646016) with the short-wavelength, low-resolution (SL)
module and the long-wavelength, low-resolution (LL) module of IRS.
The observation was carried out in IRS Staring mode, covering $\sim$ 5 to 40 $\mu$m at a resolving 
power of $\lambda/\delta\lambda$ = 60-100. We extracted and calibrated the spectrum using the 
Spectral Modeling, Analysis, and Reduction Tool (SMART) software package \citep[IRS instrument team][]{higdon04}.
More details on the data reduction
can be found in
\citet{furlan06}. 

\section{Analysis and Results}
\label{sec:results}

In this section we examine the emission of the disks detected by PACS in the Orion OB1a and OB1b association by comparing the predictions of irradiated accretion disk models to their SEDs. We estimate the stellar properties of our sample and describe the dust structure needed to explain the emission. For objects with sub-mm data, we also show our disk mass estimates.

\subsection{Spectral Energy Distributions}
\label{sec:seds}

\begin{figure*}[ht]
\epsscale{1.0}
\plotone{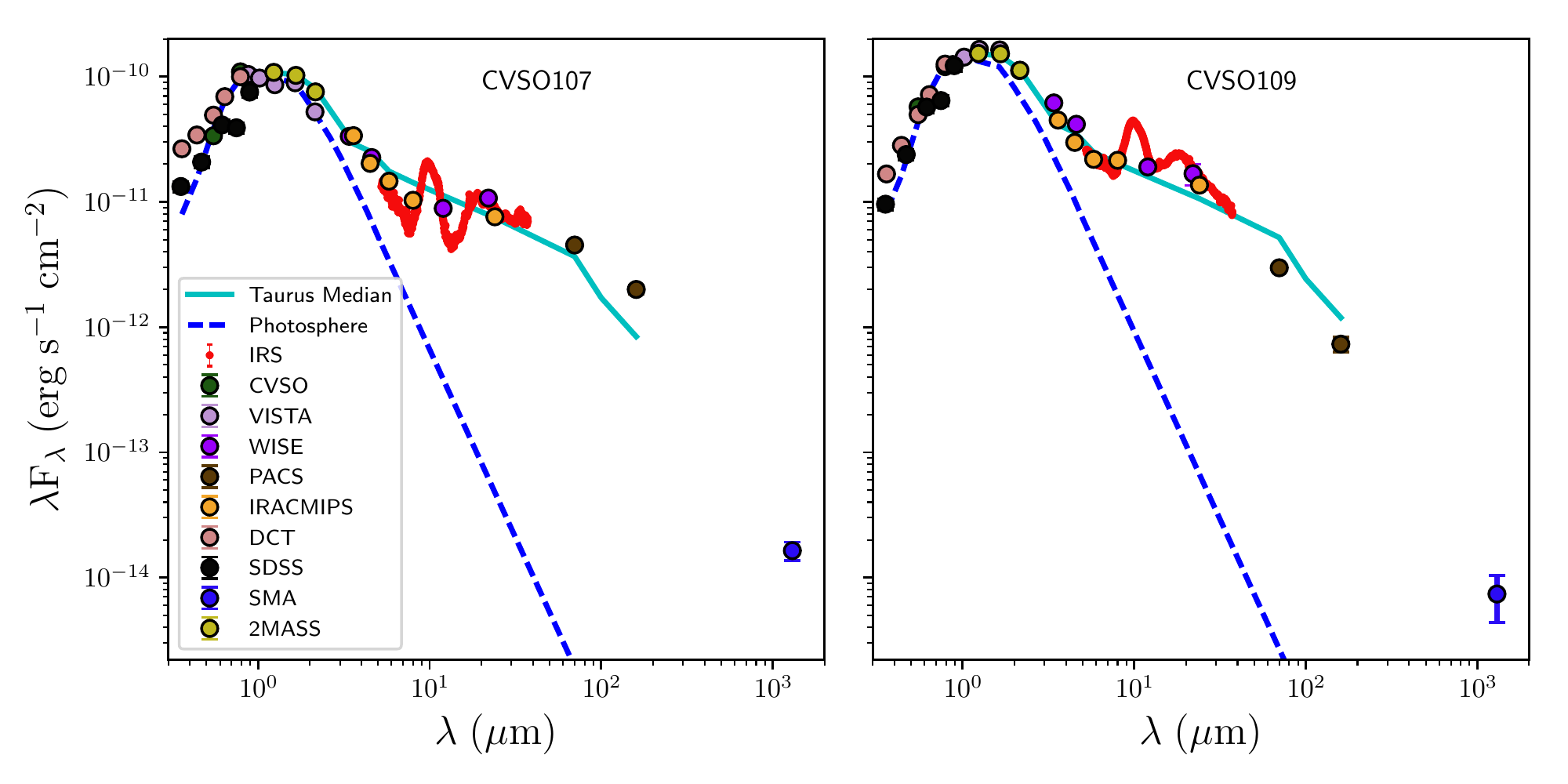}
\caption{Dereddened SEDs of the two stars in our sample that have \emph{Spitzer} IRS spectra. 
Optical data are from the CVSO \citep[][Brice\~no et al. 2018, in preparation]{briceno05, briceno07b}, the 
SDSS and the DCT, the near-IR magnitudes at J, H, and $\rm K_S$ are from the 2MASS \citep{skrutskie06}, and at Z, Y, J, H, $\rm K_S$, from VISTA \citep{petr-gotzens2011}. The mid-IR measurements, at 3.6, 4.5, 5.8, 8, and 24 $\mu$m are from our \emph{Spitzer} GO-13437 and GO-50360 programs \citep{hernandez07a}, and those at 3.4, 4.6, 12 and 22 $\mu$m are from the AllWISE Source Catalog \citep{wright2010}.
The IRS spectra are shown as solid red lines.
The dashed-blue line is the stellar photosphere (normalized to the J band of each object) of the same spectral type \citep{kh95},
and the light-blue solid line is the Taurus median from \citet{mauco16}. Error bars are typically smaller
than the symbols.}
\label{fig:sedsIRS} 
\end{figure*}

\begin{figure*}[ht]
\epsscale{1.2}
\plotone{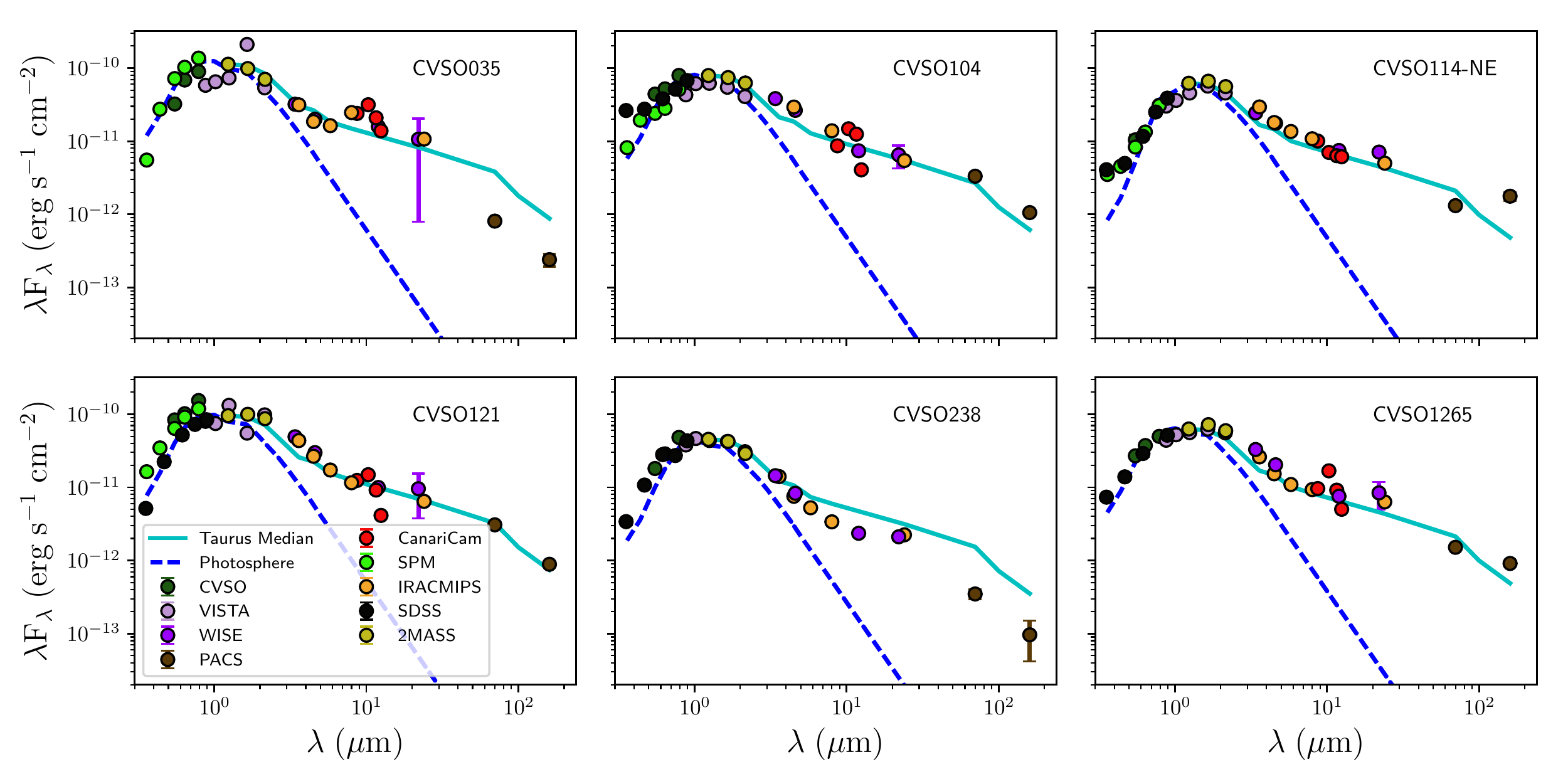}
\caption{Dereddened SEDs of stars in our sample without IRS spectra. Optical data are from the CVSO \citep[][Brice\~no et al. 2018, in preparation]{briceno05, briceno07b}, the 
SDSS and SPM, the near-IR magnitudes at J, H, and $\rm K_S$ are from the 2MASS \citep{skrutskie06}, and at Z, Y, J, H, $\rm K_S$, from VISTA \citep{petr-gotzens2011}. The mid-IR measurements at 3.6, 4.5, 5.8, 8, and 24 $\mu$m are from our \emph{Spitzer} GO-13437 and GO-50360 programs \citep{hernandez07a}, at 8.7, 10.3, 11.6, and 12.5 from CanariCam, and those at 3.4, 4.6, 12 and 22 $\mu$m are from AllWISE Source Catalog \citep{wright2010}.
Symbols and line styles are the same as in Figure~\ref{fig:sedsIRS}.}
\label{fig:seds} 
\end{figure*}

In Figure~\ref{fig:sedsIRS} we show the dereddened SEDs
of our PACS sample for sources with \emph{Spitzer} IRS spectra.
Figure~\ref{fig:seds} shows the SEDs for those
sources for which we do not have IRS data.
We adopted the Mathis reddening law \citep[$R$ = 3.1]{mathis90} 
with visual extinctions, $A_{\rm v}$, from \citet{briceno05} and Brice\~no et al. (2018, in preparation). 
The dashed line indicates the stellar photosphere (normalized to the J band of each object) of the same spectral type following \citet{kh95}. The light-blue
solid line is the Taurus median (estimated from photometric data only) for K \& M, class II stars taken from \citet{mauco16}.
All our objects exhibit excesses over the photosphere from the near-IR to millimeter wavelengths consistent with
the presence of a disk. Additionally, CVSO-35, 104, 121, and 1265 with CanariCam photometry
and especially CVSO-107 and CVSO-109 with IRS spectra
show strong 10 $\mu$m silicate emission.

CVSO-35, 104, 107, and 121 show significant variability in optical and near-IR as seen from the scatter in 
the V, $\rm R_c$, $\rm I_c$ magnitudes as well as in the J, H, and $\rm K_S$ bands taken at different epochs. 
Although most of our objects have typical full-disk excess emission beyond 20 $\mu$m,
with SED comparable to the median of Taurus,
objects like CVSO-107 and 238, and possibly
CVSO-104 and 121 seem to have 
a flux
deficit around $10-12$ $\mu$m
relative to the median, possibly the result of the first stages in the development of an inner disk gap; if so, these objects could be in the process of evolving to a transitional disk, like 
CVSO-224 \citep[see][]{espaillat2008}, which shows a deep emission deficit at wavelengths between $\sim 8$ $\mu$m and 22 $\mu$m, the telltale of an inner disk hole cleared of dust.
In the case of CVSO-114NE PACS fluxes at wavelengths longer than 10 $\mu$m look flatter than in the rest of the sample.
This star forms an apparent pair with the
star CVSO-114SW,
separated by 4.9\arcsec \citep{thanathibodee2018} 
Both components have been observed and resolved using far-UV, optical and near-IR spectroscopy as well as high angular resolution imaging;
analysis of the accretion properties indicate that
the north east (NE) component, studied here, is a CTTS, while the south west (SW) component is a WTTS \citep{thanathibodee2018}.
The SW component is only brighter than the NE component in the optical, 
so it is unlikely to have a contribution in the 
PACS range.

\subsection{Stellar and Accretion Properties}
\label{sec:stellar_para}

We estimated stellar and accretion properties of all the 
CTTS detected by PACS in Orion OB1a and OB1b reported as members in \citet{briceno05,briceno07a}, for which we
have the necessary spectra and photometry. 
Table~\ref{tab:properties} lists the results as follows:
target CVSO ID from \citet{briceno05}, spectral type, effective temperature ($T_{\rm eff}$),
visual extinction ($A_{\rm v}$), stellar mass ($M_{*}$), stellar radius ($R_{*}$), stellar luminosity ($L_{*}$), 
accretion rate ($\dot M$), stellar age, and
distance. 
To characterize the stellar properties of the sources we located them
in the HR diagram.
For this, we
estimated the luminosity of our sample
using 2MASS J
photometry,
visual extinctions and spectral types
from \citet[2018, in preparation]{briceno05}.
We used bolometric corrections and effective temperatures
from the standard table 
for 5-30 Myr old
PMS stars from \citet[]{pecaut13}.
Using these luminosities and effective temperatures we estimated 
stellar radii. We used the PMS evolutionary tracks of \citet{siess00} to obtain stellar masses. 
We assumed a distance for the OB1b association of 440 pc and distances 
of 354 pc and 368 pc for the 25 Ori and HR1833 stellar aggregates, respectively \citep[2018, in preparation]{briceno05}.
Since \citet{pecaut13} only include 3 optical magnitudes (B, V, and $\rm I_c$) we used the intrinsic colors from 
\citet{kh95}, which includes more optical magnitudes (U, B, V, $\rm R_c$, and $\rm I_c$ ), to represent the stellar photosphere.

Mass accretion rates ($\dot M$)
were estimated from the $\rm H\alpha$ line luminosity. 
The $\rm H\alpha$ luminosity was estimated
by approximating the flux of the line (F$_{\rm H\alpha}$) as EW$\rm (H\alpha$) $\times$ F$_{\rm cont}$,
where F$_{\rm cont}$ and EW$\rm (H\alpha$) are the continuum flux around the line and 
the equivalent width, respectively. In turn, we calculated F$_{\rm cont}$ from 
the R$_{\rm c}$ magnitude of each source \citep{briceno05} corrected by 
extinction, and its 
equivalent width EW(H$\alpha$)
from low resolution spectra (Brice\~no et al. 2018, in preparation.)
Finally, we used the relation between the H$\alpha$ luminosity and mass accretion rate from 
\citet{ingleby13}.
All the accretion parameters of PACS CTTS are shown in Table~\ref{tab:properties}.

Two of our sources, CVSO-107 and CVSO-109, have reported accretion rates in \citet{ingleby14}. 
They estimated $\dot{M}$ by fitting the excess in spectra taken with 
the Magellan Echellette Spectrograph (MagE)\footnote{http://www.lco.cl/telescopes-information/magellan/instruments/instruments/mage} using accretion 
shock models from \citet{calvet98}. 
Our estimate agrees with their results for CVSO-107 and is
consistent within a factor of 2 for CVSO-109. 
Differences are mainly due to uncertainties in stellar mass, radius, and 
extinction.

\begin{deluxetable*}{lccccccccc}

\tablecaption{Properties of PACS CTTS sources in the OB1a and OB1b subassociations \label{tab:properties}}
\tablehead{
\multicolumn{1}{c}{CVSO ID}  & \multicolumn{1}{c}{Spt} & \multicolumn{1}{c}{$T_{\rm eff}$} & \multicolumn{1}{c}{$A_{\rm v}$} & \multicolumn{1}{c}{$M_{*}$} & \multicolumn{1}{c}{$R_{*}$} & \multicolumn{1}{c}{$L_{*}$} &\multicolumn{1}{c}{$\dot M$} &\multicolumn{1}{c}{Age}
&\multicolumn{1}{c}{$d$} \\ \noalign{\smallskip}
\colhead{} & \colhead{}  & \colhead{(K)} & \colhead{(mag)} 
& \colhead{(\rm M$_{\odot}$)} &  \colhead{(\rm R$_{\odot}$)}  &  \colhead{(\rm L$_{\odot}$)} & \colhead{($10^{-9}$ \rm M$_{\odot}$ yr$^{-1}$)}
& \colhead{(My)} & \colhead{(pc)}
}
\startdata
CVSO-35    & K6   & 4020 & 0.3  & 0.73 & 1.67 & 0.66 & 0.87        & 3.0  & 354\\
CVSO-104   & K7   & 3970 & 0.1  & 0.67 & 1.77 & 0.67 & 5.62        & 2.6  & 440\\
CVSO-107   & K7   & 3970 & 0.4  & 0.66 & 2.07 & 0.96 & 2.89$^{a}$  & 1.6  & 440\\
CVSO-109   & M0   & 3770 & 0.0  & 0.50 & 2.66 & 1.28 & 6.71$^{a}$  & 0.6  & 440\\
CVSO-114NE & M1.5 & 3560 & 0.0  & 0.47 & 1.87 & 0.47 & 0.52$^{a}$  & 1.5  & 440\\
CVSO-121   & K6   & 4020 & 0.4  & 0.73 & 1.92 & 0.86 & 2.32        & 2.4  & 440\\
CVSO-238   & M0.6 & 3630 & 0.0  & 0.42 & 1.52 & 0.36 & 3.71        & 2.5  & 440\\
CVSO-1265  & K7   & 3970 & 0.0  & 0.70 & 1.32 & 0.39 & 1.51        & 6.7  & 368 \\
\enddata
\tablecomments{Column 1: ID following \citet{briceno05}; Column 2: spectral type;
Column 3: effective temperature; Column 4: visual extinction;
Column 5: stellar mass, Column 6: stellar radius; Column 7: stellar luminosity;
Column 8: mass accretion rate; 
Column 9: age; Column 10: distance \citep{briceno05} and Brice\~no et al. 2017 (in preparation).
$^{a} \dot M$ estimated using $\rm R_c$ magnitude from DCT. 
}
\end{deluxetable*}

\subsection{Disk Models}
\label{sec:disk models}

We used the ``D'alessio Irradiated Accretion Disk" (DIAD)
models from \citet{dalessio06}
in order to fit the SEDs of our sources.
These models
assume the disk is heated by stellar irradiation and viscous
dissipation. The viscosity is parameterized
through $\alpha$ \citep{shakura73} assuming steady accretion 
with constant $\dot{M}$.
To simulate the settling of dust, \citet{dalessio06} considered two populations of dust
grains that follow a size distribution
$\propto a^{-3.5}$; where $a$ is the radius of the grain, between 
$a_{\rm min}$ and $a_{\rm max}$ \citep{mathis77}. 
We assumed that the grains are segregated spheres \citep{pollack94}.
One population
consists of small 
($a_{\rm max} = 0.25 \mu$m) grains dominant in the upper layers of the disk 
while the other is described by larger grains in the disk mid-plane. 
The settling of dust is parametrized with the parameter
$\epsilon$, defined as
$\epsilon = \zeta_{\rm small}/\zeta_{\rm std}$, where $\zeta_{\rm small}$ is
the dust-to-gas mass ratio of 
small grains and
$\zeta_{\rm std}$ is
the sum of the assumed mass abundances of the different dust components
relative to gas i.e., $\epsilon$ describes the depletion 
(in mass) of small grains 
relative to the standard value.  
Therefore, lower values of $\epsilon$ represent more settled disks.

The main input parameters are 
the stellar properties ($M_{*}$, $R_{*}$, $L_{*}$), 
the mass accretion rate ($\dot M$),
the viscosity parameter ($\alpha$), the 
disk outer radius ($R_{\rm d}$),
the cosine of the inclination angle ($\mu$),
the maximum grain size
at the disk mid-plane (amax$_{\rm b}$),
at the disk inner edge or wall (amax$_{\rm w}$), and at the
disk upper layers (amax$_{\rm s}$),
and the the dust settling parameter $\epsilon$.

Transitional disks 
are modeled with
an optically thick outer disk
with a sharp inner edge (``wall").
Pre-transitional disks
also have a truncated outer disk
but require
optically thick material near the star, and thus 
a gap exists
between the inner and the outer thick material
\citep{espaillat07}.

We assumed a composition for the optically thick disks of amorphous silicates 
(pyroxenes) with a mass fraction relative
to gas of $\zeta_{sil}$ = of 0.004
and of carbonates in the form of graphite with  $\zeta_{graf}$ = 0.0025. 
We computed opacities for the silicates and graphite
grains with Mie theory with optical constants from \citet{dorschner95} and \citet{drainelee84}, respectively.
We also included 
H$_{2}$O ice opacities, calculated
with optical constants 
from \citet{warren84} and with $\zeta_{ice}$ = 0.002.
We considered a case
with no water ice (a mass fraction of $10^{-5}$) in the outer disk also. 
We used a maximum grain size at the disk midplane of amax$_{\rm b}$ = 1 mm.  

In pre-transitional disks,
the inner disk has a sharp inner edge (``inner wall") 
located at the dust destruction radius for silicates
grains. The emission from the inner
wall is calculated from the stellar properties, the
maximum grain size (amax$_{\rm w}^{i}$), and the temperature ($T^{i}_{\rm w}$),
assumed to be the sublimation temperature of silicate grains \citep[1400 K]{muzerolle03,dalessio06}, 
with a dust composition of pyroxenes
and $\zeta_{sil}$ = 0.004. 
The height of the inner wall ($z^{i}_{\rm w}$)
was fixed to 4 times the gas scale height. 
The stellar radiation impinges
directly onto the wall which we assume vertical.
We calculate the structure and emission
of the wall 
atmosphere following the prescriptions of \citet{dalessio04,dalessio05}.
We calculated the location and
height of the inner edge
of the outer disk (`outer wall') by
varying its radius
$R^{o}_{\rm w}$, or equivalently
its temperature $T^{o}_{\rm w}$
\citep[see][]{dalessio05,espaillat10}
to achieve the best fit to the SEDs.

In both pre-transitional and transitional disks, the gap or hole sometimes
contains a small amount of optically thin dust which
contributes to 
the 10 $\mu$m silicate emission feature.
We calculated the emission from this optically thin dust region
following \citet{calvet02}.  
The optically thin dust inside the gaps/holes is composed of amorphous silicates (olivines),
amorphous carbon and organics. Troilites, 
and crystalline silicates in the form
of enstatite and forsterite 
were only included in modeling objects with \emph{Spitzer}/IRS spectra. 
Opacities and optical constants for organics
and troilite were
adopted from \citet{pollack94} and \citet{begemann94}, respectively.   
We added organics
and troilite to the dust mixture following \citet{espaillat10} 
with $\zeta_{org}$ = 0.0041 and $\zeta_{troi}$ = 0.000768 and sublimation
temperatures of $T_{org}$ = 425 K and $T_{troi}$ = 680 K.
For the amorphous carbon we use $\zeta_{amc}$ = 0.001 and for silicates $\zeta_{sil}$ = 0.004.
The opacity for crystalline silicates is taken from \citet{sargent09}.
We did not include ice in
the optically thin region since the temperatures here are high
enough for it to be sublimated.
Opacities were calculate using Mie theory, assuming spherical grains \citep{pollack94}.
We note, however, that we do not aim to model the detail
composition of dust in this region but rather illustrate what typical 
dust compositions can reasonably describe the observed SEDs. 

\subsection{SED Modeling of (Pre)Transitional Disks}
\label{sec:modeling}

We have calculated detailed 
disk structures for 8 CTTS 
detected in our PACS 70 and 160 $\mu$m survey in the Orion
OB1a and OB1b subassociations.
We used the DIAD models 
\citep{dalessio2006}, constrained by the
mass accretion rates estimated
independently from optical spectra (\S~\ref{sec:stellar_para}).  
As a result of our modeling, all our objects turned out to be
PTD/TD, characterized by small deficits of mid-IR emission along
with strong silicate features at 10 $\mu$m. 
We found that the 
emission of the silicate bands cannot be reproduced
by the classical full disk model; instead we needed a dust distribution 
characteristic of (pre)transitional disks with optically thin dust inside their gaps/holes.  
We inferred the properties of the edge or ``wall" of the 
outer disk, the size of the cavity, 
the mass and composition of the optically thin dust 
inside the cavity and, for
those objects with millimeter photometry, we also estimated 
disk masses and radii.

We used as input for the models the stellar
properties, accretion rates and distances reported in Table~\ref{tab:properties}.
For each object we 
calculated a total of 2160 optically thick disk models and
more than 1500 optically thin dust models. All the relevant parameters we varied are listed in Table~\ref{tab:mod_var}.
We selected as the best fit the model that yielded the
minimum value of the reduced $\chi^{2}_{\rm red}$. All our objects have PTD morphologies, except CVSO-35, which is a TD.

\begin{deluxetable}{lr}

\tablecaption{Model Parameters \label{tab:mod_var}}
\tablehead{
\colhead{Parameter} & \colhead{value}
}
\startdata
cosine of inclination angle, $\mu$ ..& 0.3, 0.6, 0.9\\
\hline
\multicolumn{2}{c}{Optically Thick Outer Disk}\\ \noalign{\smallskip}
\hline
$\alpha$ .............................................& 0.01, 0.001\\ 
$\epsilon$ ..............................................& 0.01, 0.001\\
$R_{\rm d}^{o}$ (AU) ..................................& 200, 300\\
$\rm H_{2}O$ ice ...................................& 1e-5, 0.002\\
\hline
\multicolumn{2}{c}{Optically Thick Outer Wall}\\ \noalign{\smallskip}
\hline
amax$^{o}_{\rm w}$ ($\mu$m) ...........................& 0.25, 10, 1000\\
$T^{o}_{\rm w}$ (K) ....................................& 40, 80, 120, 160, 200\\
$z^{o}_{\rm w}$ (in units of $H$) ...................& 0.1-2.5, in steps of 0.1\\
$\rm H_{2}O$ ice ...................................& 1e-5, 0.002\\
\hline
\multicolumn{2}{c}{Optically Thick Inner disk}\\ \noalign{\smallskip}
\hline
amax$^{i}$ ($\mu$m) ............................& 0.25, 10\\
$\epsilon$ ..............................................& 0.1, 0.01\\
graf$_{\rm ab}$ ......................................& 0.0025, 0.25\\
$R^{i}_{\rm d}$ (AU) ..................................& 0.1, 0.15, 0.2\\
\hline
\multicolumn{2}{c}{Optically Thin Region}\\ \noalign{\smallskip}
\hline
$R_{\rm i,thin}$ (AU)  .............................& 0.1, 0.5\\
$R_{\rm o,thin}$ (AU) ........................... & 3, 5, 7, 10, 20, 30, 40, 50\\
$p$ ............................................. & 0.0, 0.3, 0.6, 0.9\\
amax$_{\rm thin}$ ($\mu$m) ........................& 0.25, 100 \\
\enddata

\end{deluxetable}

Since these are variable stars (see Figures~\ref{fig:sedsIRS} and~\ref{fig:seds}) and we have
multi-epoch optical and mid-IR photometry at different bands with very similar wavelengths, 
we included stellar variability in the estimate of the $\chi_{\rm red}^{2}$ by considering 
the weighted average between photometric bands taken at almost the same wavelength.
The following pairs of photometric bands were averaged: 
SDSS(riz)-CVSO(VRI), 2MASS(JHK)-VISTA(JHK) and 
IRAC/MIPS(3.6,4.5,24)-WISE(W1,W2,W4). 
For objects without SDSS photometry or incomplete CVSO photometry, 
we used SPM and DCT data instead. When multi-epoch photometry was available, we took the maximum difference between photometric values as the standard error used in the estimate of the $\chi_{\rm red}^{2}$. However, since our main purpose is to model the emission from the disk rather than 
stellar variability, 
we assigned a 90\% weight to data with wavelength larger than 2 $\mu$m and the remaining 10\% to the 
optical data in the final $\chi_{\rm red}^{2}$.  

\begin{figure*}[ht]
\epsscale{1.2}
\plotone{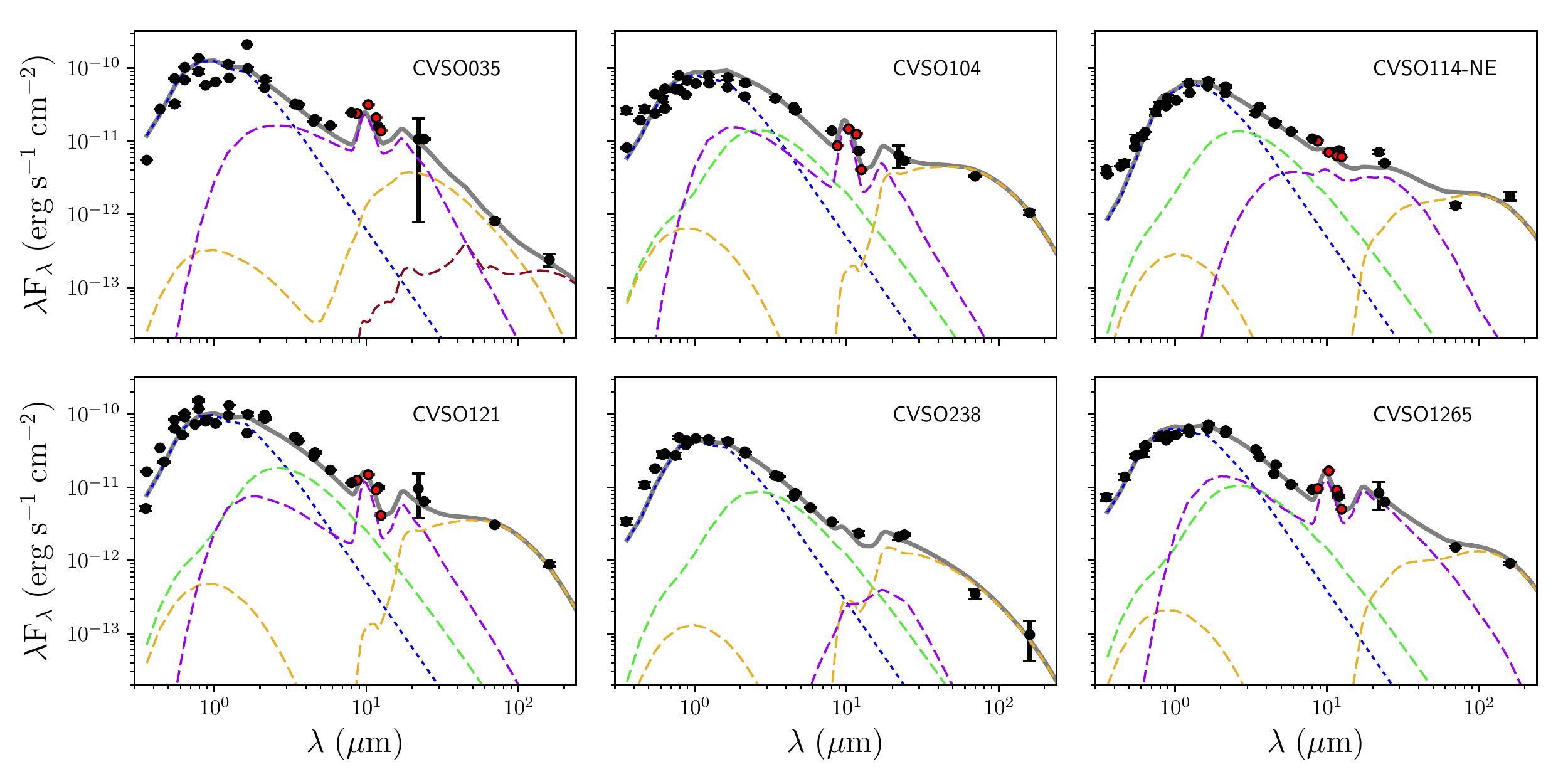}
\caption{Best-fit model (solid line) of the SED of PACS OB1a and 1b sources without IRS spectra. 
As in Figures~\ref{fig:sedsIRS} and~\ref{fig:seds}, photometry has been
dereddened following the Mathis law \citep[$R$ = 3.1, solid circles]{mathis90}. 
Dotted lines correspond to the photosphere-like fluxes using the colors of \citet{kh95} (normalized to the J band of each object). 
Dashed lines represent the model components in the following way:
inner wall+disk (green), optically thin dust region (purple), outer wall (yellow) and
outer disk (brown).
Error bars are included, but in most cases are
smaller than the symbol. The CanariCam photometry has been highlighted (red).}
\label{fig:seds_bestfit} 
\end{figure*}

\begin{figure*}[ht]
\epsscale{1}
\plotone{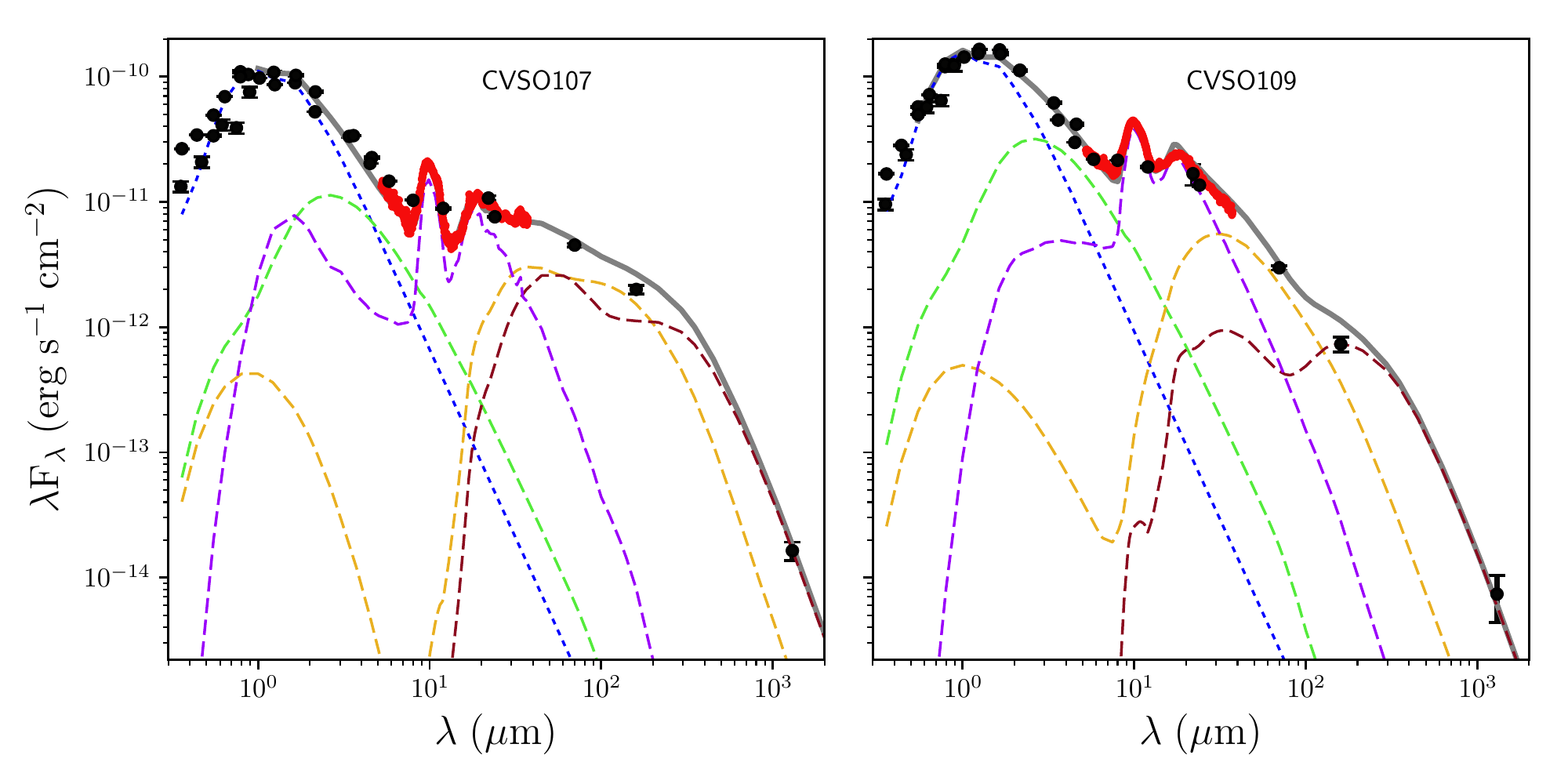}
\caption{Best-fit model (solid line) of the SED of PACS OB1a and 1b sources with IRS spectra. 
As in Figures~\ref{fig:sedsIRS} and~\ref{fig:seds}, photometry has been
dereddened following the Mathis law \citep[$R$ = 3.1, solid circles]{mathis90}. 
Line styles and colors are the same as in Figure~\ref{fig:seds_bestfit}.
IRS spectra are indicated by red lines.
}
\label{fig:seds_bestfit_IRS} 
\end{figure*}

Figure~\ref{fig:seds_bestfit} and~\ref{fig:seds_bestfit_IRS} show the SEDs of our PACS disks (solid circles)
with the resulting fit (solid lines).
The contributions of the different model
components are also shown.
The
CanariCam photometry around 10 $\mu$m as well as IRS spectra are highlighted (red).
Tables~\ref{tab:models_para} and~\ref{tab:models_othin} list the parameters of the best-fit model for each
object; in Table~\ref{tab:models_para} 
we show the outer disk properties  
and in
Table~\ref{tab:models_othin}
the properties
of the optically thin
dust region.

\begin{figure*}[ht]
\epsscale{1}
\plotone{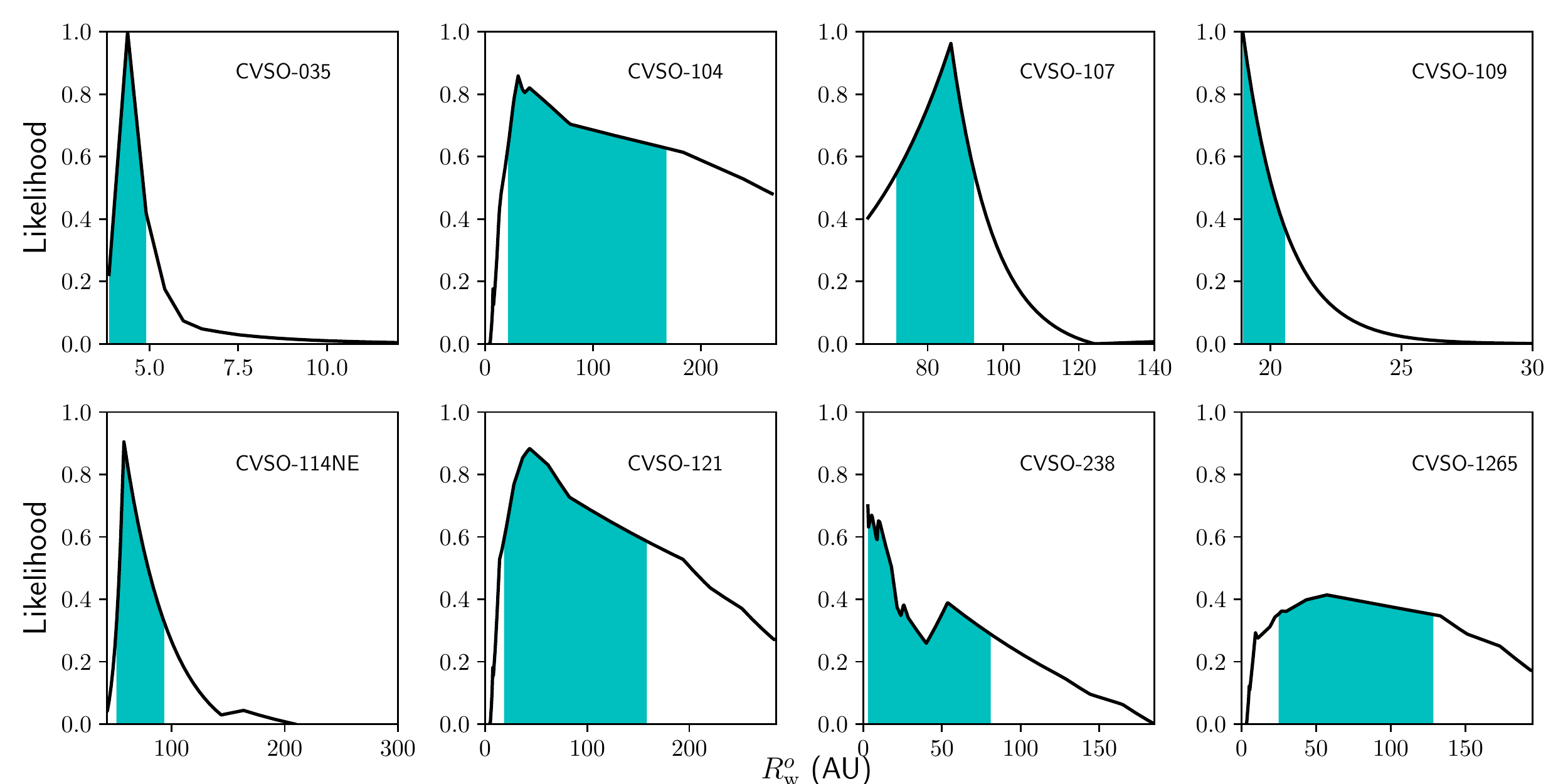}
\caption{Likelihood function, $\mathcal{L}$, vs. outer wall location, $R^{o}_{\rm w}$, for each source. 
The x axis has been restrained around the maximum peak in each case in order to better
visualize the likelihood function. Confidence intervals of $R^{o}_{\rm w}$ are shown as 
light blue shaded regions and are defined as the intervals that enclose 63\% (1-$\sigma$) of the total area
of $\mathcal{L}$. Confidence intervals are reported in Table~\ref{tab:models_para}.}
\label{fig:oRin_int} 
\end{figure*}

\begin{figure*}[ht]
\epsscale{1}
\plotone{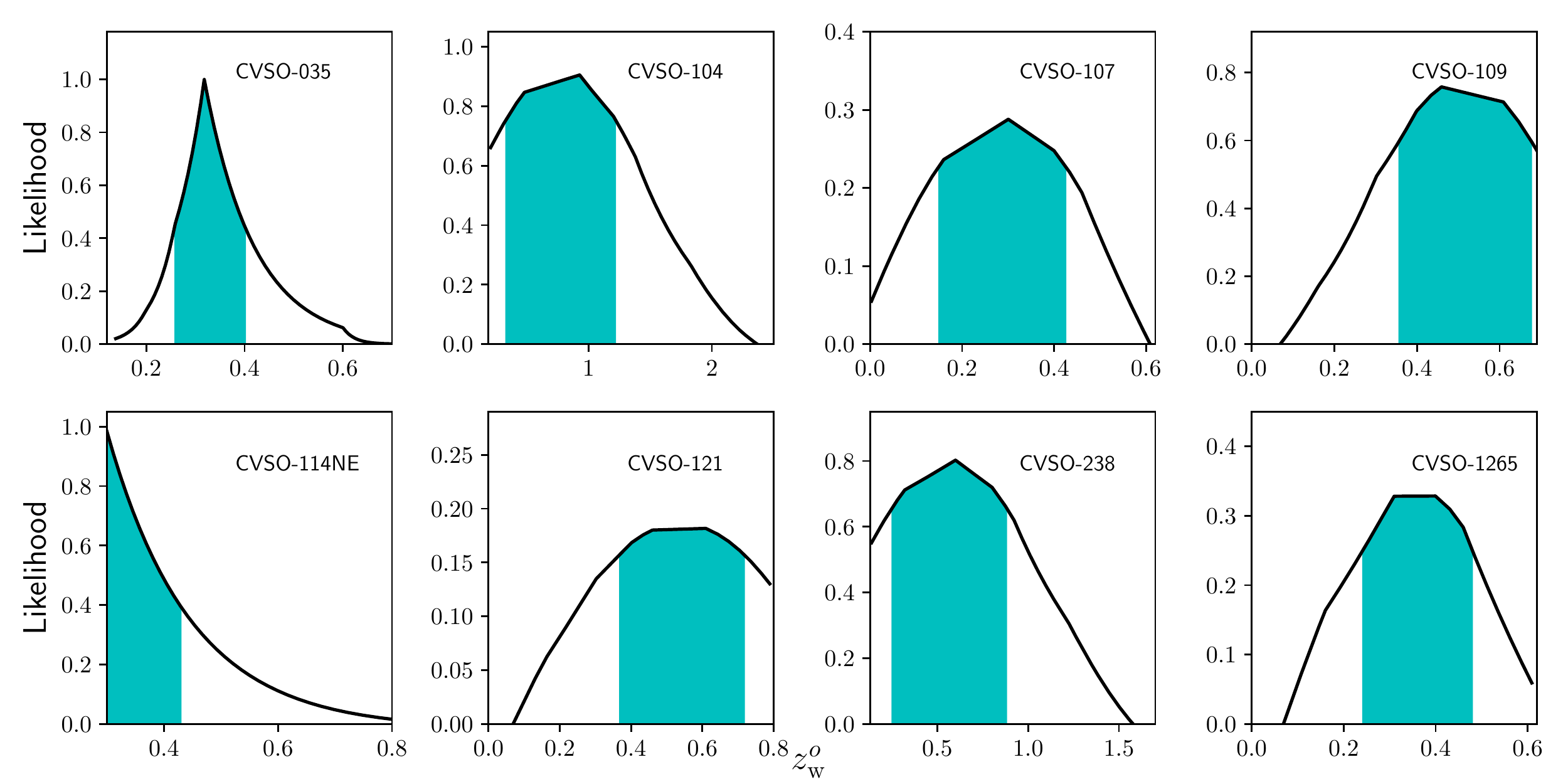}
\caption{Likelihood function, $\mathcal{L}$, vs. outer wall height, $z^{o}_{\rm w}$ (in units of the gas scale height $H$), for each source. 
The x axis has been restrained around the maximum peak in each case in order to better
visualize the likelihood function. Confidence intervals of $z^{o}_{\rm w}$ are shown as 
light blue shaded regions and are defined as the intervals that enclose 63\% (1-$\sigma$) of the total area
of $\mathcal{L}$. Confidence intervals are reported in Table~\ref{tab:models_para}.}
\label{fig:owall_int} 
\end{figure*}

As shown in Figures~\ref{fig:seds_bestfit} and~\ref{fig:seds_bestfit_IRS},
PACS fluxes are almost completely dominated by the wall of the outer disk
for most of our sources, so we
cannot constraint the properties of the outer disks for objects without mm photometry.
However, we did estimate confidence intervals for the location of the outer wall and
its height.
To set these intervals we first estimated the likelihood function, $\mathcal{L}$,
which is related to the $\chi^{2}_{\rm red}$ values through the expression
$\mathcal{L} = exp(-\chi^{2}_{\rm red}/2)$. Since $\chi^{2}_{\rm red}$ is a multidimensional function, at
every $R^{o}_{\rm w}$ or $z^{o}_{\rm w}$ we have several values of $\chi^{2}_{\rm red}$, one for each one of
the calculated models. Thus, the likelihood $\mathcal{L}$ is computed
using the minimum $\chi^{2}_{\rm red}$ value in each case.
Figures~\ref{fig:oRin_int} and~\ref{fig:owall_int} show the likelihood function for 
$R^{o}_{\rm w}$ and $z^{o}_{\rm w}$, respectively.
The confidence intervals are given as those extreme limits
at which the area below the likelihood curve maximum is 63\% (1-$\sigma$) of its total
area \citep{sivia12}.
These intervals are indicated by
light-blue shaded regions in each panel and are reported in
Table~\ref{tab:models_para}. For those cases where the best parameter falls
on one of the edges of the range of values used in the models, we
have considered these values as upper or lower limits, and they
are indicated by parenthesis instead of square brackets in
Table~\ref{tab:models_para}.

Even though we are including
the errors in the photometry and a proxy of the star variability
in the estimate of the $\chi_{\rm red}^{2}$, 
there are 
other sources of uncertainty such as the inherent
uncertainties in the distance, spectral types and
mass accretion rates. Therefore, the $\chi_{\rm red}^{2}$ should be taken only
as a guide in order to obtain the model that provides the best-fit to the
photometry for each source and not as an actual estimate of the goodness of the fit.

\subsubsection{Inner Disk and Optically Thin Region}

The inner parts of our PTD
consist of an optically thick dusty belt within the first $\sim$ 0.2 AU from the star 
and with grains that can reach 10 \um in size. The innermost edge of this ring
of optically thick material 
is located at the dust destruction radius given by the sublimation temperature of
silicate grains (1400 K).

All our objects needed the presence of a small amount of optically thin dust
inside their cavities.
Keeping the total mass fraction
of silicates relative to gas
constant, 
$\zeta_{sil} =$ 0.004, we
varied the fractional abundance
of the different silicate species
(see section~\ref{sec:disk models}) inside the gaps/holes,
the extension of the optically thin region ($R_{\rm i,thin}$, $R_{\rm o,thin}$), 
the maximum size of the dust grains (amax$_{\rm thin}$) and
the exponent of the power-law describing the dust distribution ($p$) in order to fit
the silicate feature. 
The total emission of the optically thin region was scaled to the vertical optical depth
at 10 \um ($\tau_{0}$). Table~\ref{tab:mod_var} describes the parameter space  
we used in order to model the optically thin dust inside the cavities while
Table~\ref{tab:models_othin} lists its estimated properties.

\begin{figure*}[ht]
\epsscale{0.8}
\plotone{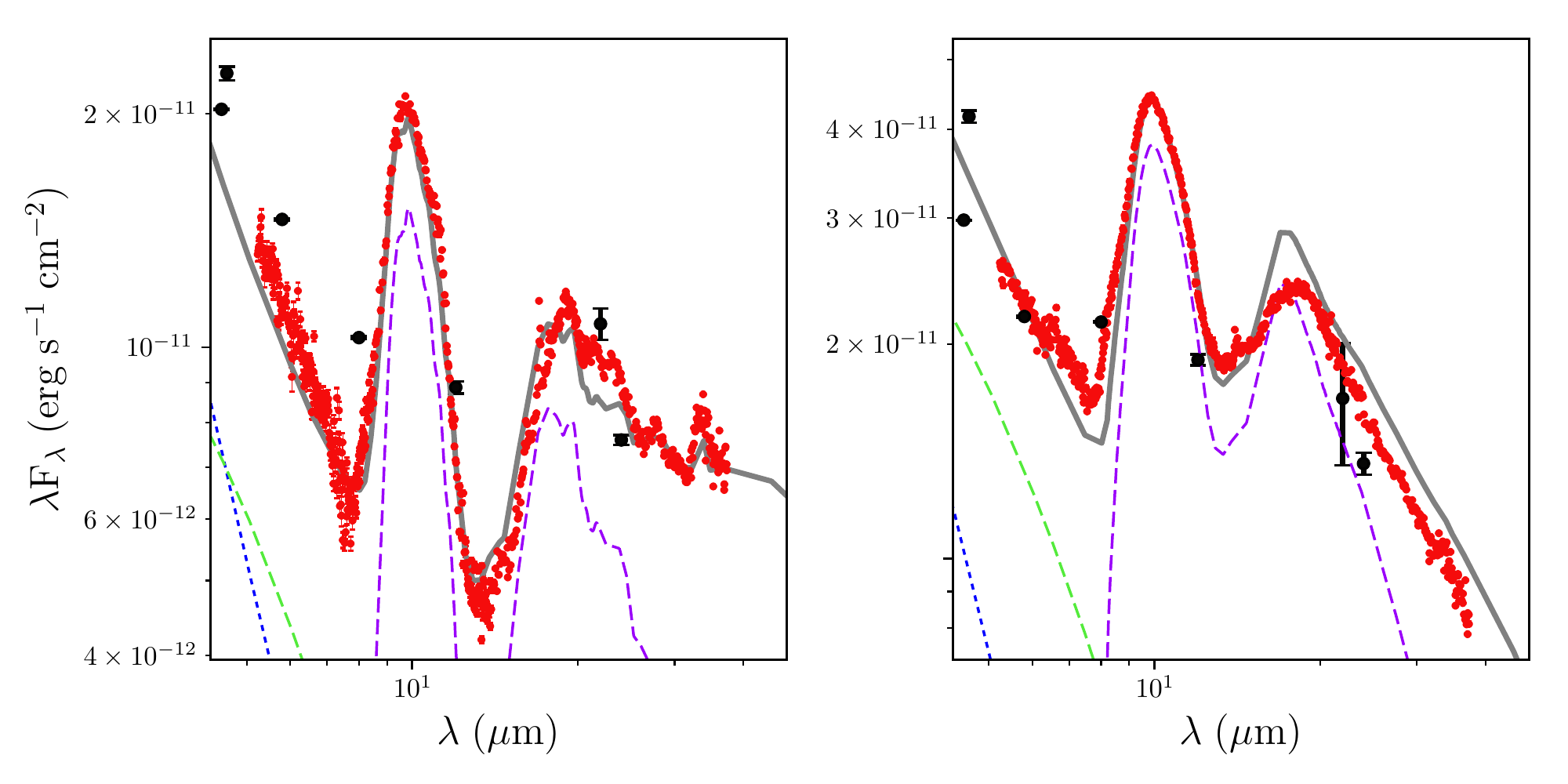}
\caption{Best-fit model (solid line) to the IRS spectra (red dots) of CVSO-107 (left) and CVSO-109 (right). 
Dotted lines correspond to the
photosphere-like fluxes using the colors of \citet{kh95} (normalized to the J band of each object). 
Dashed lines represent the model components in the following way:
inner wall+disk (green), optically thin dust region (purple). Error bars are included,
but in most cases are smaller than the symbol.
}
\label{fig:seds_bestfit_IRS_zoom} 
\end{figure*}

The silicate emission feature of all our objects with CanariCam photometry except CVSO-114NE 
can be explained with small sub micron-sized grains. 
CVSO-114NE exhibits no 10 \um silicate feature, indicating a
lack of small grains; in this case,
we found that larger grains (amax$_{\rm thin}$ = 10 $\mu$m) 
can describe the emission.
Our modeling of the silicate features
was more detailed 
for those objects with \emph{Spitzer}/IRS spectra.
The silicate feature of CVSO-109 resembles that of a pristine 
spectra, e.g with no signs of dust processing \citep{watson09}, and it is composed of dust made up 
entirely by amorphous silicates ($\sim$99\%).
CVSO-107 on the other hand, 
shows forsterite and enstatite features beyond 20 \um in its IRS spectra (e.g the 33 \um forsterite feature).
We found an optically thin dust composition consistent
with $\sim$79\% amorphous silicates, $\sim$13\% forsterite  and enstatite crystals, and $\sim$9\% organics.
Figure~\ref{fig:seds_bestfit_IRS_zoom} 
shows the fit to the IRS spectrum range
for both sources.

The 
mass of optically thin dust populating the cavities (Table~\ref{tab:models_othin}) ranges
from 7.2 x $10^{-11}$ to 1.44 x $10^{-8}$ $\msun$. 
According to
Table~\ref{tab:models_othin},
the optically thin dust
region extends to more than 50\% 
the size of the disk gap
for about half of the sample,
and about 30\% or less for the other half.
These results should be taken as an approximation 
of how much dust is required within the gaps/holes to be able
to explain the observed emission and not as a detailed prescription
of the actual spatial distribution of dust. 
High-resolution
IR interferometry 
is needed to probe the morphology of this component in detail.

\subsubsection{Outer Disk}
The disks in the sample have
a wide range of gap/hole sizes,
from $\sim$4 AU to almost 90 AU in radius. 
Large confidence intervals obtained for some objects reflects 
the need of acquiring mid-IR spectra along with (sub)mm data to better constraint cavity sizes.
We estimated outer disk properties for CVSO-107 and CVSO109, the only two sources
with SMA detections in the sample. 
The disks required low values of the viscosity
parameter (Table~\ref{tab:models_para}),
similarly to other
PTD/TD \citep{espaillat07,espaillat08a,espaillat10},
and significant degree of dust settling, $\epsilon$ $\le$ 0.01.
CVSO-107 has a disk radius of $R_{\rm d}$ = 300 AU while
CVSO-109 has a smaller disk with $R_{\rm d}$ = 200 AU. 

In CVSO-35, the only TD in the sample, 
the PACS emission is not entirely dominated by the outer wall, but seems to have a small contribution from a low-mass optically thick outer disk (See Figure~\ref{fig:seds_bestfit}).
We found a disk with $\alpha$ = 0.01, $\epsilon$ = 0.001, and $R_{\rm d}$ = 300 AU. 
However, these values should be taken as an approximation
since we do not have (sub)mm data 
for this source.

\begin{deluxetable*}{lccccccccccccccccc}
\centering
\footnotesize\setlength{\tabcolsep}{2.5pt}

\tablecaption{Outer disk properties \label{tab:models_para}}
\tablehead{
\multicolumn{1}{l}{CVSO ID} & \multicolumn{1}{c}{$\mu$} 
&  \multicolumn{1}{c}{$\rm H_{2}O$ ice}
& \multicolumn{1}{c}{} & \multicolumn{6}{c}{Optically Thick Outer Wall} 
& \multicolumn{1}{c}{} &
\multicolumn{5}{c}{Optically Thick Outer Disk} 
& \multicolumn{1}{c}{} &
\multicolumn{1}{c}{$\chi_{\rm red}^{2}$} \\ \cline{5-10} \cline{12-16} \noalign{\smallskip}
\colhead{} & \colhead{} & \colhead{} & \colhead{} & \colhead{amax$^{o}_{\rm w}$} &
\colhead{$T^{o}_{\rm w}$} & \colhead{$z^{o}_{\rm w}$} & \colhead{$(z^{o-}_{\rm w}, z^{o+}_{\rm w})$} & 
\colhead{$R^{o}_{\rm w}$} & \colhead{$R^{o-}_{\rm w}$, $R^{o+}_{\rm w}$} & \colhead{} &
\colhead{$\alpha$} & \colhead{$R^{o}_{\rm d}$} & \colhead{$\epsilon$} & \colhead{$M_{\rm dust}$} 
& \colhead{$M_{\rm disk}$} & \colhead{} &\colhead{} \\
\colhead{} & \colhead{} & \colhead{} & \colhead{} & \colhead{($\mu$m)} &
\colhead{(K)} & \colhead{} & \colhead{} & 
\colhead{(AU)} & \colhead{(AU)} & \colhead{} &
\colhead{} & \colhead{(AU)} & \colhead{} & \colhead{($\rm M_{\earth}$)} 
& \colhead{($\msun$)} & \colhead{} &\colhead{} 
}
\startdata
CVSO-35    & 0.3 & 0.002  & & 10   & 200 & 0.36 & [0.25,0.40] & 4.3  & (3.9, 4.9]     & & 0.01  & 300 & 0.001 & 3.5  & 0.0016 & & 26.77 \\ 
CVSO-104   & 0.6 & 0.002  & & 0.25 & 120 & 0.9  & [0.32,1.22] & 26.5 & [21.5, 168.7] & & --    & --  & --    & --    &  --    & & 1.50  \\  
CVSO-107   & 0.9 & 1e-5   & & 0.25 & 80  & 0.3  & [0.15,0.43] & 86.2 & [71.7, 92.6]  & & 0.001 & 300 & 0.01  & 116.8  & 0.054  & & 22.01 \\ 
CVSO-109   & 0.6 & 1e-5   & & 10   & 120 & 0.45 & [0.35,0.68) & 18.9 & (19.0, 20.6]  & & 0.006 & 200 & 0.001 & 29.0  & 0.0134 & & 10.06 \\ 
CVSO-114NE & 0.3 & 1e-5   & & 0.25 & 80  & 0.3  & (0.3,0.43]  & 57.9 & [51.4, 93.6]  & & --    & --  & --    & --    & --     & & 0.47  \\ 
CVSO-121   & 0.3 & 0.002  & & 0.25 & 120 & 0.54 & [0.37,0.72] & 27.8 & [19.1, 159.1] & & --    & --  & --    & --    & --     & & 0.56  \\ 
CVSO-238   & 0.9 & 0.002  & & 0.25 & 160 & 0.6  & [0.25,0.88] & 8.8  & (2.9, 81.0]   & & --    & --  & --    & --    & --     & & 1.03  \\ 
CVSO-1265  & 0.3 & 1e-5   & & 0.25 & 80  & 0.3  & [0.24,0.48] & 57.0 & [24.9, 129.0]  & & --    & --  & --    & --    & --     & & 0.50 \\ 
\enddata
\tablecomments{Column 1: ID following \citet{briceno05}; Column 2: cosine of inclination angle;
Column 3: $\rm H_{2}O$ ice abundance;
Column4: maximum grain size of the outer wall; Column 5: outer wall temperature; Column 6: outer wall height (in units of 
the gas scale height $H$); Column 7: outer wall height confidence intervals (in units of 
the gas scale height $H$); 
Column 8: location of the outer wall; Column 9: confidence intervals of the location of the outer wall;
Column 10: outer disk viscosity; Column 11: outer disk radius; Column 12: degree of dust settling of the outer disk;
Column 13: outer disk dust mass; Column 14: outer disk total mass; Column 15: $\chi_{\rm red}^{2}$.
Total disk mass estimates are
assuming a dust-to-gas mass ratio similar to that of the ISM.}
\end{deluxetable*}

\begin{deluxetable*}{lcccccccccccccccc}
\centering
\footnotesize\setlength{\tabcolsep}{2.5pt}

\tablewidth{0pt}
\tablecaption{Inner disk and optically thin dust region properties \label{tab:models_othin}}
\tablehead{
\multicolumn{1}{l}{CVSO ID} & \multicolumn{3}{c}{Optically Thick Inner Disk} &  \multicolumn{1}{c}{} &
\multicolumn{12}{c}{Optically Thin Dust Region} \\
\cline{2-4} \cline{6-17} \noalign{\smallskip}
\colhead{} & \colhead{amax$^{i}_{\rm w}$} & \colhead{$R^{i}_{\rm w}$} & 
\colhead{$R^{i}_{\rm d}$} & \colhead{} &
\colhead{$R_{\rm i,thin}$} & \colhead{$R_{\rm o,thin}$} & \colhead{amax$_{\rm thin}$} & \colhead{$\tau_{0}$} &  \colhead{$p$} &
\colhead{sil} & \colhead{org} & \colhead{amc} & \colhead{troi} & 
\colhead{fors} & \colhead{enst} & \colhead{$M_{\rm dust,thin}$}\\
\colhead{} & \colhead{($\mu$m)} & \colhead{(AU)} & 
\colhead{(AU)} & \colhead{} &
\colhead{(AU)} & \colhead{(AU)} & \colhead{($\mu$m)} & \colhead{} &  \colhead{} &
\colhead{\%} & \colhead{\%} & \colhead{\%} & \colhead{\%} & 
\colhead{\%} & \colhead{\%} & \colhead{$10^{-9}\msun$}
}
\startdata
CVSO-35      & -- & --   & --    &  & 0.01 & 5.0  & 0.25 & 0.08  & 0.0 & $\sim$57 & $\sim$29   & $\sim$14  & 0.0  & $<$0.2 & $<$0.2 & 0.37 \\
CVSO-104     & 10 & 0.09 & 0.10  &  & 0.1  & 20.0 & 0.25 & 0.08  & 0.6 & $\sim$85 & 0.0     & $\sim$14  & 0.0  & $<$0.2 & $<$0.2 & 4.72 \\
CVSO-107     & 10 & 0.1  & --    &  & 0.1  & 50.0 & 0.25 & 0.04  & 0.2 & $\sim$79 & $\sim$9    & 0.0    & 0.0  & $\sim$10   & $\sim$3   & 14.46\\ 
CVSO-109     & 10 & 0.11 & 0.15  &  & 0.5  & 5.0  & 0.25 & 0.1   & 0.0 & $\sim$99 & 0.0     & 0.0    & 0.0  & $<$0.2 & $<$0.2 & 1.65 \\
CVSO-114NE   & 10 & 0.07 & 0.2   &  & 0.2  & 10   & 10   & 0.1   & 0.0 & $\sim$23 & $\sim$71   & $\sim$6   & 0.0  & $<$0.2 & $<$0.2 & 2.86 \\
CVSO-121     & 10 & 0.09 & 0.10  &  & 0.1  & 20   & 0.25 & 0.035 & 0.2 & $\sim$85 & 0.0     & $\sim$14  & 0.0  & $<$0.2 & $<$0.2 & 1.65 \\    
CVSO-238     & 10 & 0.06 & 0.15  &  & 0.5  & 3.0  & 0.25 & 0.03  & 0.9 & $\sim$23 & $\sim$71   & $\sim$6   & 0.0  & $<$0.2 & $<$0.2 & 0.072\\    
CVSO-1265    & 10 & 0.06 & 0.20  &  & 0.1  & 30   & 0.25 & 0.055 & 0.0 & $\sim$72 & 0.0     & $\sim$27  & 0.0  & $<$0.2 & $<$0.2 & 7.90 \\ 
\enddata
\tablecomments{Column 1: ID following \citet{briceno05}; 
Column 2: maximum grain size in the inner wall;
Column 3: dust destruction radius; Column 4: inner disk radius; 
Column 5 and 6: extension of the optically thin dust region;
Column 7: maximum grain size of optically thin dust; Column 8: vertical optical depth at 10 $\mu$m;
Column 9: power-law exponent of the distribution of optically thin dust, Column 10-15: 
percentages of dust species in the optically thin region, Column 16: optically thin dust mass.}
\end{deluxetable*}

\subsubsection{Disk Masses}

The dust mass of the best-fit models for
objects in which we probe the outer disk
are given in Table~\ref{tab:models_para}.
The corresponding total disk mass, 
with a dust-to-gas mass ratio
$\zeta = 0.0065$ (the sum of our assumed
abundances), is also given.
We note, however, that the disk mass for CVSO-35 should be taken with caution since we need (sub)mm data
to properly constraint this parameter. 

We compared the masses obtained through detailed modeling of the SEDs with 
disk masses estimated using the SMA fluxes at 1.3 mm of CVSO-107 and CVSO-109 and assuming optically thin
emission.  
Following \citet{hildebrand83}: 
\begin{equation}
 M_{\rm dust} = \frac{F_{\nu}d^{2}}{\kappa_{\nu}B_{\nu}(T_{\rm dust})},
 \label{opticallythinmass}
\end{equation}
\noindent
where $F_{\nu}$ is the submillimeter flux
at 1.3 mm,
$d$ is the source distance,
and
$T_{\rm dust}$ is a characteristic dust temperature assumed to be 
the median for Taurus disks \citep[][$T$ = 20 K]{andrews05}, $B_{\nu}$ is the
Plack function at $T_{\rm dust}$ and $\kappa_{\nu}$ is the dust grain opacity
taken as 0.1 $\rm cm^{2}g^{-1}$ at 1000 GHz using an opacity power-law index
of $\beta = 1$ \citep{beckwith90}. 
Using the above equation and a dust-to-gas mass ratio of 0.01 we computed $M_{\rm disk}$ of 0.0087 $\msun$ and 0.0039 $\msun$
for CVSO-107 and CVSO-109, respectively. 
These masses are a factor of $\sim$6 and $\sim$3 lower than the masses obtained from detailed modeling (Table~\ref{tab:models_para}).
The difference is due in part to the
difference in opacities (our opacity is a factor of 2 lower than the $\kappa_{\nu}$ used in eq.(\ref{opticallythinmass})) and dust to gas mass ratios,
to the assumption of constant temperature, 
and also due to the contribution from 
hotter, optically thick disk regions to the 1.3 mm
flux, not included in eq.(\ref{opticallythinmass}).
We stress
that we used a consistent opacity law for each one of our objects, which depends on
the mix of materials assumed in our disk models (silicates,
graphite, water, etc.), their abundances, and their grain size
distributions, rather than 
the simplified approach of a single grain opacity in an isothermal disk to estimate disk masses reported in Table~\ref{tab:models_para}.

\section{Discussion}
\label{sec:discussion}

\subsection{Why these objects fail to be described as full disks?}

\begin{figure}[ht]
\epsscale{1}
\plotone{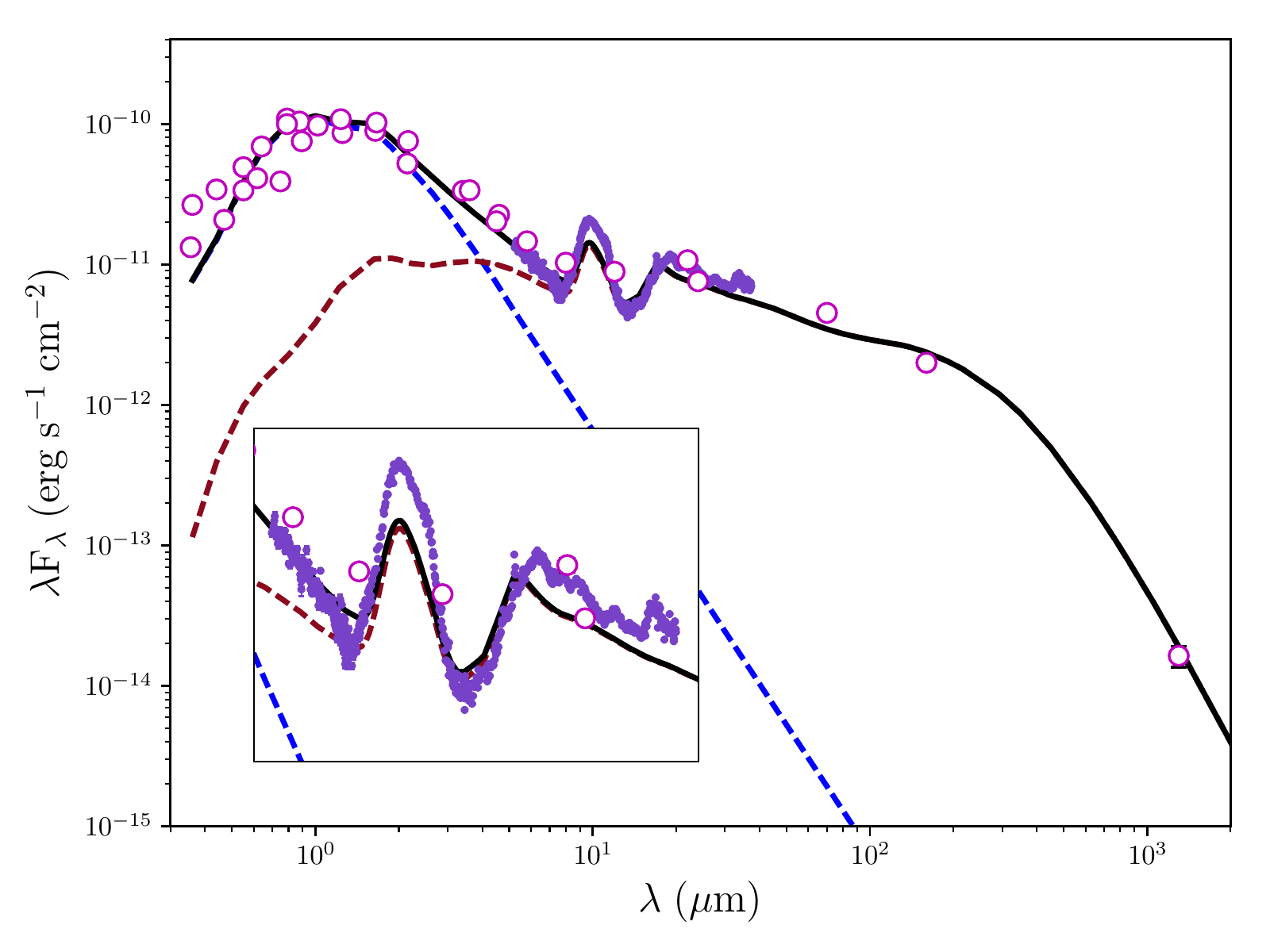}
\caption{Best-fit to the SED of CVSO-107 considering full disk models. 
Magenta open circles indicate photometric 
data while the IRS spectra
is shown as purple dots. The solid black line represents the best-fit. 
The model consists of a stellar photosphere (blue-dashed line) and a disk (brown-dashed line).
Note how full disks are unable to 
generate enough near- and mid-IR emission especially at the silicate features at 10 \um and 20 $\mu$m.}
\label{fig:CVSO107_full_disk} 
\end{figure}

\begin{figure}[ht]
\epsscale{1}
\plotone{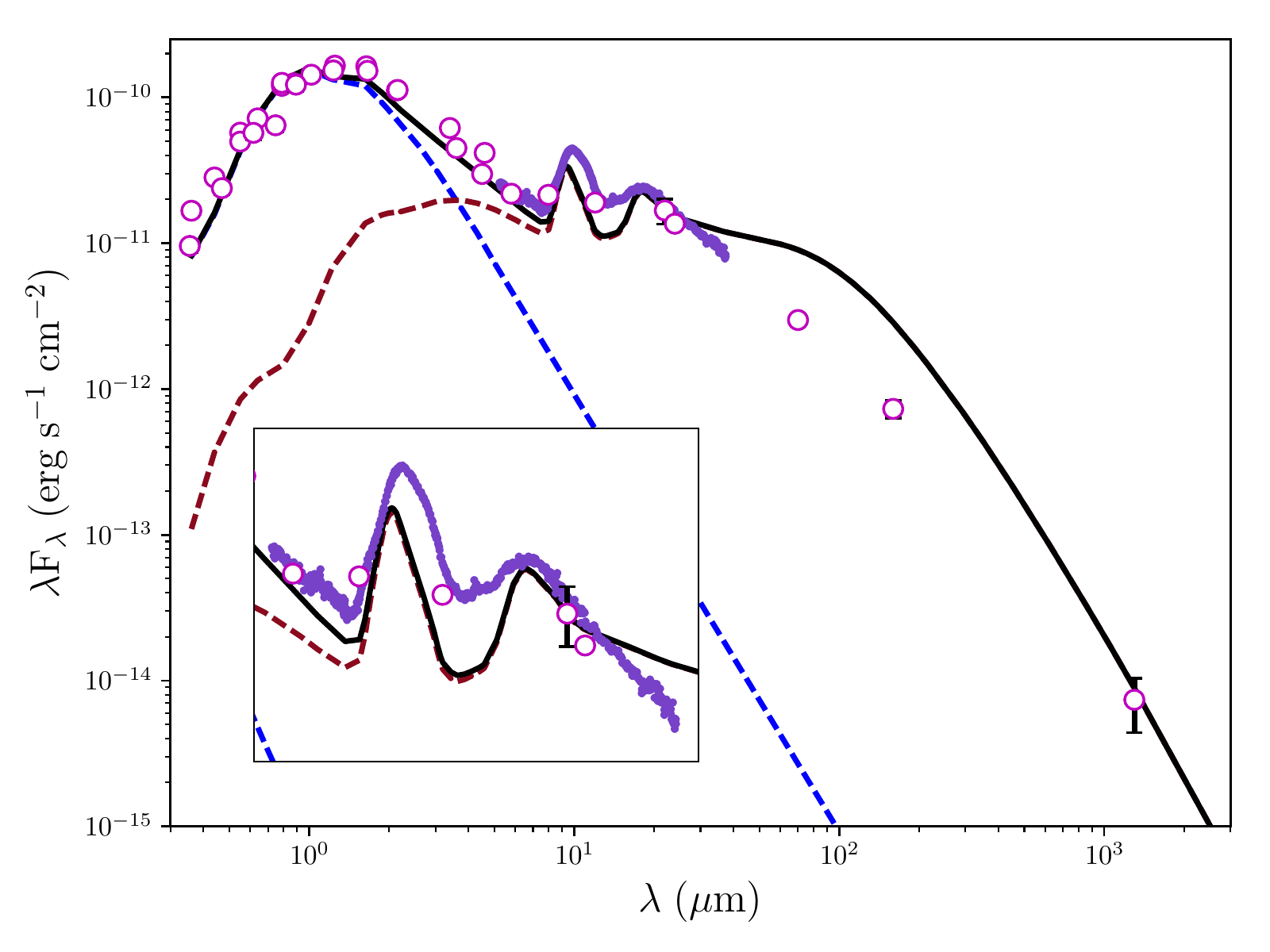}
\caption{Best-fit to the SED of CVSO-109 considering full disk models. 
Magenta open circles indicate photometric 
data while the IRS spectra
is shown as purple dots. The solid black line represents the best-fit. 
The model consists of a stellar photosphere (blue-dashed line) and a disk (brown-dashed line).
Note how full disks are unable to 
generate enough near- and mid-IR emission especially at the silicate features at 10 \um and 20 $\mu$m. The model also overestimates the emission beyond 26 $\mu$m.}
\label{fig:CVSO109_full_disk} 
\end{figure}

In order to model our disk sample, we included both full and truncated disk models as priors,
and the best fit was obtained
with PTD/TD.
We first note the small decrease of mid-IR emission 
in the SEDs of some of our objects
compared to the median of CTTS in Taurus (Figure~\ref{fig:seds}, light-blue line). 
This points to a lack of optically thick material similar to those found in known 
pre-transitional disks \citep{espaillat08a,espaillat10}.
More importantly, full disk models do not appear to produce high enough IR emission at 10 \um 
to explain the observed silicate features, even in objects with IRS spectra as CVSO-107 and CVSO-109.
Figures~\ref{fig:CVSO107_full_disk} and \ref{fig:CVSO109_full_disk} show the best-fit to the SEDs of CVSO-107 and CVSO-109, respectively, considering full disk models.
As shown, the model underestimates near- and mid-IR excesses, especially at the silicates bands
at 10 \um and 20 $\mu$m (see figures insets). For CVSO-109 the model also overestimates the emission beyond 26 $\mu$m.   
A viable way to generate high enough emission at the silicate bands and 
a small decrease in the mid-IR emission  
is by including optically thin dust inside gaps.
Even in the case of CVSO-238, which does not have
photometry nor spectra around 10 $\mu$m, we were unable to model its PACS photometry
along with its WISE 22 \um and IRAC/MIPS 24 \um photometry with a full disk. 

Given the intermediate-advanced age of this region, is not unreasonable 
to think that these objects may have experienced significant 
evolution over time. 
Based on our disk mass estimates for CVSO-107 and CVSO-109, 
we can assume that early on in their lives
there was probably enough material to form multiple planets. If this is indeed the case, 
then these planets might be responsible for the radial structures--characteristic of disks with gaps and holes--observed here.
However, there are other possible
explanations.
For instance, magnetized disks without planets \citep{flock15}, or fast pebble growth near condensation fronts \citep{zhang15}
may create structures
in disks. 
In these cases, rings in disks can be the precursors of planets rather than the cause \citep[e.g.,][]{carrasco16}.
The clear presence of near-IR excesses in the SEDs of all our stars,
along with significant mass accretion rates, indicate
that there is still gas and dust in the inner regions,
and therefore these objects can be classified as (pre-)transitional disks based on the observational data currently available for these sources. We note, however, that high-resolution IR interferometric observations are still needed in order to confirm the morphology of these disks.

\subsection{Implications for dust evolution}

Investigating dust evolution in the outer disk requires 
far-IR observations of a significant sample of TTS 
spanning the first several Myr 
in the lives of low-mass stars. Since by 5 Myr only about 20\% of the stars
still retain their inner disks \citep{hernandez07b}, 
disk populations at intermediate ages 
are essential
to link currently disk properties with evolutionary processes.

As shown in Figure~\ref{fig:seds_bestfit}, and discussed in the last section, the SED modeling of our sources 
indicates that these are PTD/TD.
Moreover, while most (pre)transitional disk studies have focused on 
young star-forming regions of less than 3 Myr, few 
have been done to address the physical mechanisms responsible for the existence 
of (pre)transitional disks in older regions.

The intermediate-advanced age of our sources ($\sim$4-10 Myr) poses two possible evolutionary scenarios: 
1) that these stars had full disks until recently, and have now
become PTD/TD or 2) that the PTD/TD appearance is long-lasting, in which case we are actually looking at ``mature" PTD/TD systems. This last argument is reinforced by the fact that all our PACS detections turn out to be PTD/TD. This challenges the current understanding of disk evolution
where pre-transitional disks morphology is thought to be a transient stage.
Moreover, dust evolution models struggle to find
viable ways to explain inner disk survival 
for long periods of time. 

One possible mechanism that seem to explain pre-transitional disk appearance is dust filtration induced by the presence of embedded planets \citep[e.g.][]{paardekooper06,rice06,fouchet07,zhu12,espaillat14,pinilla15}. Since dust and gas in disks
are not perfectly coupled, gas drag forces dust grains to drift towards a pressure maximum \citep{weidenschilling77,johansen14}.
This filtration process will lead to discontinuous grain populations in the radial direction, with small grains in the inner disk and larger grains outward.
Recently, \citet{pinilla16} studied partial filtration of dust particles to explain   
the survival of the inner disks in PTD by combining hydrodynamical 
simulations of planet-disk interactions with dust evolution models.
According to them, in systems forming low-mass planets ($<$ 1 M$_{\rm jup}$), 
the micron-sized particles ($\le$ 1 $\mu$m)
are not perfectly trapped at the outer edge
of the planet-gap, but in constant movement through the gap via turbulent diffusion. 
This partial filtration of grains
supports a constant replenishment 
of small dust from the outer to the inner disk. As
a consequence, the near-IR excess can remain for up to 5 Myr of evolution 
and the SED morphology remains almost identical. 
They concluded that the near-IR excess that
characterizes pre-transitional
disks is not necessarily an evolutionary effect, but
depends on the type of planets sculpting the disks.
Our sources are at the upper end of ages
studied by \citet{pinilla16} (they only considered 1 and 5 Myr old disks), 
nonetheless, 
if this effect remains for older disks, 
which is possible since these disks are still accreting, 
then, 
\citet{pinilla16}'s results 
suggest that
the disks studied here could
be forming low-mass planets.

Some of our targets
exhibit strong silicate emission. 
The silicate feature at 10 \um carries vital information 
of the sub-micron grains left over inside the gaps and holes
of PTD/TD. 
In particular, it carries
information on dust
processing in
the inner disk through
the presence of crystalline
material. 
For the two sources with
IRS spectra,  
we were 
able to estimate the general composition of dust grains 
producing the silicate characteristic
emission. 
CVSO-109 has no signs of crystalline silicates 
while CVSO-107 exhibits some degree of dust processing 
with small amounts of enstatite and forsterite crystals.
This variety 
of dust properties for objects of the same region
is hard to explain and might indicate some correlation 
between the processes that create the optically thin dust with
disk local conditions, e.g. density and temperature, over time. 
The presence or absence of crystals inside disk cavities
sets important constraints to the density and temperature profiles
of the small dust (coupled to the gas) that will probably end up forming planets
and thus, setting their properties.
 
Total disk masses of the two targets with SMA fluxes are greater than 10 $M_{\rm jup}$, 
the minimum mass solar nebula \citep{weidenschilling77}. However, 
these masses are estimated assuming a dust-to-gas mass ratio of 0.0065.
If one considers 
larger dust-to-gas mass ratios, which are expected in older star-forming regions,
these values can easily drop below a few Jupiter masses. 
Moreover, dust disk masses for these objects are small and well below the minimum mass of solids needed
to form the planets in our Solar System \citep{weidenschilling77}.
This is consistent with previous studies in other star forming regions indicating that disks older than 5 Myr lack sufficient dust to form 
giant planet cores and therefore, time scales for giant planet formation must be quite short \citep{carpenter14,barenfeld16,barenfeld17}.\\\\

\section{Summary and Conclusions}
\label{sec:conclusion}

We  present 
Herschel PACS fluxes at 70 $\mu$m and 160 $\mu$m 
and CanariCam 10 $\mu$m
photometry of
8 CTTS in the Orion OB1a and OB1b sub-associations. We combined the
Herschel data with optical UBVRI, near and mid-IR, and sub-mm photometry, and mid-IR spectra from \textit{Spitzer},
when available, to construct
the SEDs of these sources, 
which we modeled with
irradiated accretion disk models \citep{dalessio06}.
Our main conclusions are as follows:

\begin{enumerate}
\item The best fit models to the SEDs of the targets indicate that all are PTD/TD, with some amount of optically thin dust
inside their cavities. PACS photometry was particularly useful to characterize the inner edge of the outer disks. Full disk models cannot produce enough emission at 10 \um to explain the
CanariCam photometry or the \emph{Spitzer} IRS spectra.

\item The IRS spectra of CVSO-107 and CVSO-109 can be explained with small 
grains mostly composed of amorphous silicates. The silicate feature of CVSO-109 resembles
that of a pristine spectrum with no signs of dust grain processing. In contrast, the 
IRS spectrum of CVSO-107 
is better described with the presence of  
enstatite and forsterite crystals in its optically thin dust mixture.

\item The presence of near-IR excess in the SEDs 
of our
4-10 Myr PTD sample
may point to low-mass ($<$1 $M_{\rm jup}$) planet formation.
According to \citet{pinilla16},
the survival and maintenance of the inner disk 
could be explained 
by partial filtration of dust, 
in which the micron-sized grains 
pass through the gap, supporting a constant replenishment 
of dust from the outer to the inner disk.

\item Our inferred dust disk masses, $M_{\rm dust}$, are less than the minimum mass of solids needed to form the planets in our Solar System. 
This is consistent with previous studies on disk populations older than 5 Myr, giving support to the scenario of short timescales for giant planet formation.

\end{enumerate}

\acknowledgements
We thank the referee for the helpful comments
that improved the content and presentation of
this work. We also thank the observers who were
involved in acquiring one or more
sets of observations.

K.M. acknowledges a
scholarship from CONACYT.

CCE and DMF were supported by the National Science Foundation under Grant No. AST-1455042.

These results made use of the Discovery Channel Telescope at Lowell Observatory, supported by Discovery Communications, Inc., Boston University, the University of Maryland, the University of Toledo and Northern Arizona University.
The Submillimeter Array is a joint project between the Smithsonian Astrophysical Observatory and the Academia Sinica Institute of Astronomy and Astrophysics and is funded by the Smithsonian Institution and the Academia Sinica.
The authors wish to recognize and acknowledge the very significant cultural role and reverence that the summit of Mauna Kea has always had within the indigenous Hawaiian community.  We are most fortunate to have the opportunity to conduct observations from this mountain.

Funding for SDSS-III has been provided by the Alfred P. Sloan Foundation, the Participating Institutions, the National Science Foundation, and the U.S. Department of Energy Office of Science. The SDSS-III web site is http://www.sdss3.org/.
SDSS-III is managed by the Astrophysical Research Consortium for the Participating Institutions of the SDSS-III Collaboration including the University of Arizona, the Brazilian Participation Group, Brookhaven National Laboratory, Carnegie Mellon University, University of Florida, the French Participation Group, the German Participation Group, Harvard University, the Instituto de Astrofisica de Canarias, the Michigan State/Notre Dame/JINA Participation Group, Johns Hopkins University, Lawrence Berkeley National Laboratory, Max Planck Institute for Astrophysics, Max Planck Institute for Extraterrestrial Physics, New Mexico State University, New York University, Ohio State University, Pennsylvania State University, University of Portsmouth, Princeton University, the Spanish Participation Group, University of Tokyo, University of Utah, Vanderbilt University, University of Virginia, University of Washington, and Yale University.

This research has made use of the NASA/IPAC Infrared Science Archive, which is operated by the Jet Propulsion Laboratory, California Institute of Technology, under contract with the National Aeronautics and Space Administration.

This publication makes use of data products from the Two Micron All Sky Survey, which is a joint project of the University of Massachusetts and the Infrared Processing and Analysis Center/California Institute of Technology, funded by the National Aeronautics and Space Administration and the National Science Foundation.

J.B.P. acknowledges UNAM-PAPIIT grant number IN110816, and to UNAM’s
DGAPA-PASPA Sabbatical program.

J.H. acknowledges UNAM-PAPIIT grant number IA103017.

\software{Astropy \citep{astropy13}, Matplotlib \citep{matplotlib05}, HIPE \citep{ott2010}, RedCan \citep{omaira13}, IRAF \citep{iraf99}, MIRIAD (http://www.cfa.harvard.edu/∼cqi/mircook.html), SMART \citep{higdon04}}

\facilities{Herschel(PACS),GTC(CanariCam),SMA,OANSPM:0.8m,DCT(LMI),Spitzer(IRS),WISE,Sloan,CTIO:2MASS}

\bibliography{MaucoLib,MyLibrary}

\end{document}